\documentclass[12pt]{article}
\begin{document}
\noindent
\textbf{Covariant  Hamiltonian Dynamics with Negative
Energy States}
\vskip 1.0 truecm

\centerline{M. De Sanctis $^{a,~b}$}
\vskip 1.0 truecm
\noindent
\textit{$^a$ Departamento de F\'isica, Universidad Nacional de Colombia,
Bogot\'a D. C., Colombia.  }

\noindent
\textit{$^b$ INFN sez. di Roma, P.le A. Moro 2, 00185 Roma, Italy. }

\noindent
e-mail :  mdesanctis@unal.edu.co  and   maurizio.desanctis@roma1.infn.it

\vskip 1.0 truecm
\begin{abstract}
\noindent
A relativistic quantum mechanics  is studied for bound hadronic systems
in the framework of the Point Form Relativistic Hamiltonian Dynamics. Negative 
energy states are introduced taking into account the restrictions imposed
by a correct definition of the Poincar\'e group generators. 
We obtain nonpathological, 
manifestly covariant wave equations that  dynamically contain 
the contributions of the negative energy states. 
Auxiliary negative energy states
are also introduced, specially for studying the interactions of the hadronic systems with 
external probes.
 
\vskip 0.25 truecm
\noindent
PACS number(s): 11.30.Cp, 24.10.Jv, 03.65.Pm
\end{abstract}

\vskip 1.0 truecm
\noindent 
\textbf{1. Introduction }

\vskip 0.5 truecm
\noindent
The study of hadronic  few-body systems in terms of constituent particles represents
a very important tool for the determination of  their intrinsic properties and
their interactions with external probes.
We use the term constituent particle to mean a system that transforms 
as an irreducible representation of the Poincar\'e group with definite 
mass, spin and internal symmetry quantum number.

\noindent
In this respect we recall that
historically the investigation started with the study of light
nuclei in terms
of \textit{constituent} nucleons. Later, many efforts have been devoted to
the study of the hadrons in terms of \textit{constituent} quarks.
For both cases nonrelativistic models were  initially considered, later on, relativistic
corrections were added, improving  the reproduction of the experimental data.

\noindent
Nowadays, the construction of  hadronic \textit{covariant} constituent models 
must be considered  strictly necessary for an accurate description of these systems
and for the study of their interactions with electroweak probes.
In particular, the present work, that represents a generalization of the 
Relativistic Hamiltonian Dynamics, is focussed on the study of  quark
(and quark-di\-quark) models, but its formal developments can be also applied
to the study of few-body nuclear systems. 

\vskip 0.5 truecm
\noindent
As for the relativistic covariant quark models, they 
represent \textit{effective} models that should be related
to quantum chromodynamics (QCD), incorporating in a nonperturbative way
its symmetries and dynamical features. On the other hand, the nuclear models
rely on a phenomenological meson exchange theory.

\vskip 0.5 truecm
\noindent
\textbf{1.1 General remarks}
\vskip 0.5 truecm
\noindent
From a theoretical point of wiew, we note that, for building few-body relativistic
models, two slightly different
approaches can be followed.

\noindent
The first one, is denominated Relativistic Hamiltonian Dynamics (RHD). 
The second one is represented by the Relativistic Wave Equations (RWE).
\vskip 0.5 truecm
\noindent 
The aim of the first approach, that is
the RHD,  is to satisfy the Poincar\'e Group commutation rules by defining
the generators of that group in terms of the constituent particle operators.

\noindent
We now syntetically recall some technical aspects related to RHD.
In the case of free particles the total generators of the Poincar\'e group are given
by the sum of the single particle generators. The main problem of RHD, not found 
in the nonrelativistic case, consists in fulfilling the Poincar\'e group commutation rules
when the interaction is introduced in the generators. 

\noindent 
In this respect,
RHD can be formulated in different ways, three of which have been introduced 
in the pionieristic work by Dirac [1]. They are called
the Instant Form (IF), the Front Form (FF) and the Point Form (PF).

\noindent
The most relevant difference 
among them is represented by the way in which the generators depend on the interaction.

\noindent
In the IF both the Hamiltonian, i.e. the time translation generator,
and the boost generators are modified (with respect to the free case) by the interaction
[2-4].

\noindent
In the FF, linear combinations of the components of the four-vectors 
are considered and, in consequence,
not standard Lorentz transformations must be introduced. In this form of RHD,
the interaction modifies
pieces of both the boost and of the four-momentum [3].

\noindent
Finally, in the PF [5-7], that is the scheeme adopted in the present work,
the interaction modifies the total four-momentum, i.e. the
Hamiltonian and the three-momentum of the system, that are given not only
by the sum of the four-momenta of the costituents, but also receive
a contribution that depends on the interaction operator.
On the other hand, the boost is left 
free from interaction.
Due to this last property, PF RHD has been defined as manifestly covariant [5].
This point will be analyzed in the next subsect.1.2, considering the definition
of manifest covariance adopted in the present work.

\vskip 0.5 truecm
\noindent
Theoretically, the relevant aspect of RHD is that 
its quantum mechanical properties are 
well defined, in analogy to nonrelativistic case.

\noindent
All the generators of the Poincar\'e algebra are represented by 
\textit{hermitic} operators so that the corresponding transformations are performed
by \textit{unitary} operators, satisfying  Poincar\'e covariance and allowing, 
at the same time, to use
the standard quantum mechanical procedures for the calculation of physical observables.
The interaction operator is given by a \textit{quasipotential} that is, in general,
momentum dependent.

\noindent
Note that some aspects that are typical of quantum field theories have been 
completely excluded in RHD:
in particular, the possibility of
creating or destroying  particles and
the presence of negative energy states in the interaction amplitudes.
In this work we shall analyze and overcome this second problem.
 
\vskip 0.5 truecm
\noindent
The quark models based on RHD reproduce some general hadronic features related to QCD,
like colour global symmetry, confinement and isospin invariance 
for the \textit{u d} quark sector.

\noindent
Furthermore,
in all the three forms of RHD
very encouraging quantitative calculations have been done and
are still in progress both for the hadronic
spectra and for the electroweak form factors [8]. 

\noindent
On the other hand, the \textit{dynamics} of these models
is intrinsically phenomenological and partially unsatisfactory, 
because, as discussed above, some fundamental aspects of QCD considered as field theory,
are not taken into account by standard RHD. 
Moreover, the parameters that appear in the
mass operator of these models are usually \textit{fitted} to the experimental 
data in order to reproduce the hadronic spectra.

\noindent
The electroweak form factors are calculated in the relativistic impulse (or spectator)
approximation by using a current operator that, in the case of the 
electromagnetic interaction,  is not dynamically conserved.
The discrepancies that remain between the theoretical calculations  and the
form factor experimental data are usually cured
by inserting phenomenological quark form factors and/or vector meson exchange factors
at the quark interaction vertex.

\vskip 0.5 truecm
\noindent
The second possible approach for the relativistic study of  few-body bound
hadronic systems, is represented by the use of integro-differencial
three-dimensional RWE. 

\noindent
We leave aside from our discussion the four-dimensional Bethe-Salpeter equation 
because, if it is not reduced to a three-dimensional form, 
its formalism cannot be easily interpreted in a physical way and also the practical
solution of the equation
presents serious difficulties.

\noindent
In general, the procedure to obtain the RWE (for two-body interacting systems)
starts from an underlying field theory.
Historically, quantum electrodynamics was considered. 

\noindent
From the perturbative
expansion of the field theory, a  set of relevant
Feynman graphs   is selected,
excluding, in general, the contributions due to the poles of the bosonic propagators.
Usually, the \textit {box} and (in some cases) the 
\textit{crossed box} graphs are taken. 
The infinite series of these graphs,  
denoted  respectively as \textit{ladder and crossed ladder},
is \textit{summed up} by means of a 
Lippman-Schwinger equation for the scattering matrix, in which  a suitable
Green function is  used.
Finally, from that equation, an integro-differencial, three-dimensional RWE
for the bound states is derived. 

\noindent
The origin of the differences among the various RWE is due to which Feynman graphs 
are selected and to the approximations done to sum them up in a three-dimensional form.
The procedure outlined above establishes
the connection between the RWE relativistic  model
and the underlying field theory.
Generally, the solutions of the RWE cannot be 
interpreted as standard  quantum mechanical wave functions.
Also, discussing some specific examples we shall highlight some difficulties of the RWE 
approach.

\vskip 0.5 truecm
\noindent
Historically, RWE have been introduced to study electromagnetically bound systems,
like atoms and positronium, taking quantum electrodynamics as the fundamental theory.
Later, they have been applyed to the study of few-body nuclear systems. 
In this case it was used
a phenomenological interaction operator that represents the exchange of  pions 
and heavier mesons. 

\vskip 0.5 truecm
\noindent
The simplest case of RWE is represented by the  positive energy states
Black\-en\-becler-Sugar equation [9]. 
In the remainder of this section we shall not 
further discuss this case that is not specifically relevant for the main objective
of this work. 
The Black\-en\-becler-Sugar 
equation will be considered again in sect.4
(as a special example with no negative energy states) showing that
it can be directly interpreted by means of the standard PF RHD formalism
and re-written in a manifestly covariant way.

\noindent
Considering the Dirac-like equation [10],
we recall that, in this case,  a theoretical difficulty was found
because this equation presents 
an unphysical pole for a vanishing value of the total mass 
of the system. As a consequence, a correct normalization of 
the wave function is not possible. This problem has been analyzed and
denoted as \textit{continuum dissolution}  and \textit{cockroach nest}
in refs. [11] and [12], respectively.
An analysis of this problem will be also performed
in the present work by according to the requirements of PF RHD.

\noindent
The Gross equation [13] represents a relevant example of a manifestly
covariant approach for the study of interacting systems. 
It has been obtained by means of a covariant 
procedure starting from the \textit{box} and \textit{crossed box} Feynman graphs.
Even though also in this equation a pole for a vanishing total mass is present,
this singularity is easily removed requiring  the (phenomenological) kernel
to vanish faster than the total mass. In any case the solutions of the equation are
normalizable if the total mass is not equal to zero.
The Gross equation  is not manifestly symmetric, as such, 
under particle interchange, but it has been symmetrized in all its applications.
Within this approach, the  mass of the bound system is obtained as a pole 
\textit{below threshold} of the interacting scattering matrix.

\noindent
We also consider the so-called Breit equation in the form  
originally given for the study
of electromagnetically bound systems [14]. This equation
is obtained from the four-dimensional Bethe-Salpeter equation approximating
the electromagnetic interaction with an instantaneous quasipotential.
Only \textit{box} Feynman graphs are taken. 
This equation  can be directly 
interpreted in the scheeme of RHD that will be developed in the present work. 

\noindent
Finally, the Mandelzweig-Wallace  integro-differential equation [11] 
is structurally similar to the previous one, but also includes,
in the eikonal approximation,  the \textit{crossed box}
graphs, improving the correspondence with the underlying field theory
and obtaining the so called \textit{one-body limit} 
when the mass of one particle is set equal to infinity.
This  equation can be also written in a covariant form [15, 16].

\noindent
However, a standard definition of the Poincar\'e group generators in terms of  
PF RHD quantum mechanical operators is not directly feasible within this model. 
Such problem arises because the Mandelzweig-Wallace   equation is
\textit{not} an eigenvalue equation for the mass of the system.
In other words, the mass operator is not defined explicitly.

\noindent
More recently, the Mandelzweig-Wallace formalism has been revised and denoted 
as \textit{equal time} reduction.
In a clever work [17] concerning electron scattering on the deuteron, considered as a
two-nucleon relativistic bound system, that equation has been written 
in a Hamiltonian form and
the boost generator is constructed, with some
approximations, in the instant form RHD.

\noindent
With respect to this problem, in sect.4 we shall show that, 
with some handling, an equation equivalent to the Mandelzweig-Wallace one  
can be written in the framework of 
our PF RHD model, otaining \textit{exact} relativistic covariance.
By means of this procedure we shall introduce the \textit{auxiliary states}
that represent another original aspect of our work. 
\vskip 0.5 truecm
\noindent
\textbf{1.2 Motivations and structure of the work}
\vskip 0.5 truecm
\noindent
Having discussed the theoretical framework of PF RHD and RWE, we introduce the motivations
of the present study. 

\noindent
The long term objective of the relativistic quark model investigation
would be the construction of a
covariant Hamiltonian model for the study of the hadronic systems. This model should
reproduce, with a limited number of parameters, the hadronic spectra and the response
of the hadronic particles to electroweak probes, that is elastic and inelastic form
factors, Compton scattering amplitude, etc..

\noindent
Relativistic covariance, possibly in manifest form, should represent the formal
framework of the model.

\noindent
As for the \textit{dynamics}, we point out that QCD is assumed to be the \textit{ultimate}
physical description for these system.
For this reason, the covariant Hamiltonian model should represent
a solvable approximation of QCD in the sense that its (generalized) wave equation 
should be able to \textit{sum up} the relevant QCD graphs, allowing to treat 
perturbatively other effects not included in the sum.

\noindent
The connection with QCD should allow to relate the parameters of the model with those
of QCD. In particular, we refer to the effective quark masses. 
Also, the quark-(anti)quark  interaction should be possibly derived from QCD.

\noindent
These objectives are, obviously, ambitious and lie beyond the present understanding 
of nonperturbative  aspects of field theories.
Deep and long investigations are required. 
However,
the present work, as a preliminary step, can help to solve some specific theoretical
problems, also improving the understanding of the hadronic phenomenology. 

\vskip 0.5 truecm
\noindent
As for the main objective of the present work, 
that is focussed on the study of spin $1/2$, identical, interacting quarks,
we want to include, with some approximations,  the negative energy states in PF RHD, 
in order to represent the dynamical
relativistic effects that have not been taken into account by standard RHD models.
We recall that in   quantum field theories, like QCD, when the perturbative 
series is considered, the negative energy states
appear  in the propagators of the intermediate particles.
For this reason, the procedure of the present work can help to improve 
the understanding of the link between RHD and the underlying field theory.
Also in the case of few-body nuclear systems, the
discussion about RWE shows that the contributions of the negative
energy states are considered necessary for constructing a dynamically 
consistent model. 

\vskip 0.5 truecm
\noindent
Technically, the insertion of negative energy states 
in the mass operator of the model, is obtained, in sects.2 and 3, 
by generalizing the PF RHD construction of the Poincar\'e group generators, fulfilling,
also in presence of negative energy states, the commutation rules of the algebra.
As shown in sect.3, this procedure is possible if
all the particles of the state have the \textit{same energy sign}. These states represent
the \textit{dynamical states} of our generalized PF RHD. Such condition is strictly
necessary in order to define the four-velocity operator $V^\mu$, that, in turn,
is essential to introduce
the total four-momentum of the system $P^\mu$.

\noindent
Our wave equation, written in sect.4, is a coupled equation that 
involves positive
and negative energy \textit{dynamical states} and explicitly satisfies charge 
conjugation invariance.

\vskip 0.5 truecm
\noindent 
Another original result of this work is the introduction of the \textit{auxiliary states} 
that represent the states in which particles with different energy signs are present.
For this reason (as discussed before) they do not directly participate in the dynamics 
of the model.
The \textit{auxiliary states} are covariantly defined, 
by means of  the interaction operator, in eqs.(4.14b) and (4.18b) for two and three-body
systems, respectively.

\noindent
In this way, the definition of the \textit{auxiliary states} does not require 
the introduction of new parameters.
Their expression is derived by using RWE formalism as a \textit{link} with field theory.

\vskip 0.5 truecm
\noindent
In order to understand in more detail the physical meaning of the auxiliary states,
we recall that, in perturbative expansions,
the \textit{negative energy} terms of the fermionic propagators  
give rise to the so-called
\textit{z-graphs}, corresponding to  intermediate states with one quark
and a \textit{quark-antiquark pair} [17-19].
The contributions of such states to the electroweak currents have been extensively
studied in the nonrelativistic limit, within a constituent chiral quark model.
In particular, 
the two-body (or exchange) electromagnetic four-currents have been
derived [18], obtaining 
significant contributions  for the inelastic transition amplitudes 
of all nucleon resonances.

\noindent
In the same context it has been also shown that the exchange contributions
are necessary to satisfy Partial Conservation of Axial Current (PCAC) condition
for the weak interactions, also leading to a possible interpretation 
of the missing nucleon spin as angular momentum carried by the nonvalence
degrees of freedom of the nucleon [19].

\noindent
Note that, while the electromagnetic exchange currents could be in part also
derived by means of a minimal coupling substitution in the nonrelativistic Hamiltonian,
on the other hand the derivation of the weak (axial) exchange currents strictly
requires the use of the \textit{z-graphs}.

\noindent
For the  reasons discussed above,
in our generalized PF RHD
the \textit{auxiliary states}, that represent the pair terms in effective way,
are expected to give relevant contributions to the interactions
of the hadronic system with external probes. In subsequent works it will be studied at 
numerical level their effect on electroweak form factors. 

\vskip 0.5 truecm
\noindent
Furthermore, as for the possible relevance of both 
\textit{dynamical} and \textit{auxiliary} negative energy states for 
the electromagnetic interactions of hadronic systems,
we point out that Compton scattering amplitude, if studied by means of the
Feynmam graphs of quantum electrodynamics, 
requires the presence of the negative energy states
in the fermionic propagators, in order to obtain the correct expansion 
up to the second order in photon energy [20].

\noindent
For this reason, photon scattering on nucleons (considered as bound quark states)
can represent a very helpful tool 
for the study of some nontrivial aspects of the hadronic dynamics that should be 
reproduced by RHD quark models.

\vskip 0.5 truecm
\noindent
As for the strong interaction of the quarks, the negative energy states appear in the
covariant wave equation of the model, given in eq.(4.6), at two levels. 
First, as discussed previously,
we have the \textit{dynamical} negative energy states.
Furthermore, the connection established with RWE and, in turn, with the underlying
field theory, gives rise to \textit{quadratic terms} in the effective interaction 
operators of eqs.(4.13b) and (4.17b) for two and three-body systems, respectively.
These terms reproduce, in the wave equation, the effects
due to  intermediate states (of the scattering matrix expansion)
with different energy signs.
By means of the same arguments used for the \textit{auxiliary states}, 
the \textit{quadratic terms}, representing the \textit{z-graphs} [17],
can be interpreted as contributions of \textit{quark-antiquark pairs}.
 
\noindent
Note that (as in the case of the \textit{auxilary states})
these (extra) \textit{quadratic terms} do not introduce new parameters, being related
to the linear terms of the quasipotential.
The effects of the quadratic terms on the constituent interaction
have been studied in the context of few-body
nuclear physics [13, 21]. In the case of quark models, the small value of the quark
masses can give rise to highly nonrelativistic effects that were not present in 
the interaction of the nuclear systems.

\vskip 0.5 truecm
\noindent
The kinetic term of our wave equation given in eq.(4.6) is similar to that of the Breit
equation of ref.[14], where a perturbative technique has been also studied for the
case of positronium, that is bound by the electromagnetic interaction.
Taking into account that the properties 
of the wave equation are strictly related to
the structure of the interaction operator, we note that in  the present case, 
that is quark models, such operator is not directly known and should 
be determined taking into account  the underlying field theory.

\noindent
The total effective interaction (linear + quadratic terms)
is in any case highly momentum dependent
and a carefull formal and numerical examination of the wave equation
should be performed to highlight 
its properties in connection with nonperturbative hadronic phenomenology.

\vskip 0.5 truecm
\noindent
Another objective of this work is to write the wave equations of our generalized PF RHD
in a \textit{ manifestly covariant} way. We \textit{define} here as 
\textit{ manifest covariance} the property of an equation of being written in terms of
quantities that (a) transform as tensors under Lorentz transformations and
(b) are not related to a specific reference frame.

\noindent
In this sense, the Bethe-Salpeter and Gross equations are \textit{ manifestly covariant}.
On the other hand, the wave equations of PF RHD, written in the standard formalism,
do not fully satisfy the requirement (b) 
due to the use of the zero-momentum (rest) frame 
for the definition of the velocity states [5].

\noindent
However, it is well-known that, if a theory is really covariant, with some handling
it can be written in a  \textit{ manifestly covariant} form, as it will be done for
PF RHD in the present work.

\vskip 0.5 truecm
\noindent
Obviously, \textit{ manifest covariance}  only represents a formal property
of the equations and does not lead, as such, to  improve the knowledge of the
hadronic dynamics.  

\noindent
Technically, \textit{ manifest covariance} is obtained by means of
(a) the definition of the projection states of eq.(3.9), where the three-momenta
of $N-1$ particles and spatial part of the four-velocity of the system are selected 
as spatial variables, and (b) the choice of the normalization of the wave function given 
in eq.(4.5), leading as result to a
great clarification the formalism of PF RHD.
Note that the structure of the covariant integration in eq.(4.6) is similar to that 
originally introduced by the Gross equation [13].

\vskip 0.5 truecm
\noindent
We recall that
the methods of PF RHD allow to define a  consistent framework for 
the study of the electroweak 
interactions of the hadronic systems (elastic and inelastic form factors),
specially by introducing as a starting point
the so called relativistic impulse (or spectator) approximation [5].
Our \textit{manifestly covariant} formalism also allows to calculate in a 
much more direct and clear way
the matrix elements of the electroweak current in that
approximation, as done in ref. [22] for a model with only positive energy states. 
Furthermore, in this way it is possible to introduce
a minimal coupling procedure  to derive a conserved electromagnetic current [23]
for a model containing positive energy states.
\vskip 0.5 truecm
\noindent
The formal developments of the paper are organized as follows. 

\vskip 0.5 truecm
\noindent
In sect.2, we define the
Poincar\'e Group  generators for the case of \textit{free particles} also introducing 
the negative energy states. 
This objective is reached considering that in the three-momentum Lorentz transformation
(and, in general, in the Lorentz boost operators), the relevant parameter is the product
$\lambda {\bf v}$, introduced in eq.(2.2b), being $\lambda$
the energy sign of the particle and  ${\bf v}$ the spatial part 
of the four-velocity boost parameter.

\noindent
We also study the connection between the  state representation 
of PF RHD with that given by the standard Dirac equation spinors, which is currently
used in the developments of quantum field theory.
In particular we show the equivalence of the use of the RHD Wigner rotations 
with  standard Dirac boosts for the calculation of the relevant 
matrix elements of the model.

\vskip 0.5 truecm
\noindent
In sect.3, we construct the generators of the Poincar\'e algebra 
for \textit{interacting particles} with negative energy states,
verifying that the commutation rules of the Poincar\'e Group are still fulfilled.
To this aim, we introduce the definition of \textit{dynamical states } and 
\textit{auxiliary states}. Only the former enter in the wave equation of the model.
We also introduce the  projection states  of eq.(3.9). These states  depend on the
spatial variables that are used to obtain the manifestly covariant form of the model.
Furthermore, the interaction operator is written in terms
of Dirac spinors and matrices  highlighting its covariant character and allowing
to derive its expression from field theories.

\vskip 0.5 truecm
\noindent
In subsect.4.1, we write in eq.(4.6) the wave equation of the model 
in a manifestly covariant way.
We show that
relativistic covariance, realized by the Poincar\'e Group commutation rules,
as a dynamical consequence,
automatically avoids, in our formalism, the continuum dissolution disease.

\noindent
In subsect.4.2 we use the techniques of the RWE 
(referring to the Mandelzweig-Wallace model)
to establish a \textit{link} with field theories.
In particular we introduce, starting from the same interaction operator,
both the \textit{auxiliary states} in eqs.(4.14b) and (4.18b),
and the \textit{quadratic terms} in eqs.(4.13b) and (4.17b) 
for the effective quasipotential of the wave equation. 

\vskip 0.5 truecm
\noindent
In the appendix we also show explicitly that the interaction operator is invariant
under Lorentz transformations. 

\vskip 1.0 truecm
\noindent
\textbf{2. Poincare group transfomations for single particle states}
\vskip 0.5 truecm
\noindent
As a starting point we recall the form of a Lorentz transformation (boost)
of a four-vector \(f^{\mu}=(f^0,{\bf f})\). 
In all this work we consider canonical boosts.  
The boosted four-vector \( f_b^{\mu}\) has the form
$$f^0_b=v^0f^0+{\bf v f}  \eqno{(2.1a)}  $$
$$ {{\bf f}}_b= {\bf f} +{\bf v}({\bf v f} 
\frac{1}{v^0+1} +f^0) \eqno{(2.1b)}$$
These equations can be summarized as follows
$$ f^{\mu}_b = L^{\mu}_{~\nu}( {\bf v}) f^{\nu} \eqno{(2.1c)}$$
In the previous equations we have introduced 
the relative four-velocity $ v^\mu=(v^0,{\bf v}) $ 
that satisfies $v_\mu v^\mu =1$ and, in consequence
$$v^0= v^0({\bf v})=[1+{\bf{v}}^2]^{1/2}\eqno(2.1d)$$

\noindent
The well-known relations
${\bf v}=\gamma {\bf u}$ and
$v^0=\gamma=[1-{\bf u}^2]^{-1/2}$ easily connect the four-velocity
to  the standard   \textit{physical velocity} $\bf u$
of the initial reference frame measured from the boosted one.

\noindent
We now consider the Lorentz transformation for the on-shell four-momentum 
of a single particle.
In the quantum-mechanical model that will be studied in the following, the particle three-momentum
${\bf p}$ will be used as spatial variable. In order to study also negative energy states,
we introduce the notation
$p^{\mu}(\lambda,{\bf p})=(\lambda\epsilon({\bf p}), {\bf p}) $
being $\lambda$ the energy sign of the state and
$$\epsilon({\bf p})=[{\bf p}^2+m^2]^{1/2}$$
the absolute value of the energy;
also, $m$ represents the mass of the particle.
We enphasize that in our model the  free energy of a particle
$p^0(\lambda,{\bf p})$ is always considered as a function of the
energy sign $\lambda$ and of the three-momentum ${\bf p}$ of the state.
According to eqs.(2.1 a,b) the Lorentz transformation is
$$ p^0_b(\lambda,{\bf p};{\bf v})= v^0 \lambda\epsilon({\bf p})+{\bf v p} =
 \lambda\epsilon({\bf p}_b({\bf p};\lambda {\bf v})) \eqno{(2.2a)  }$$
$$ {\bf p}_b(\lambda,{\bf p};{\bf v})= {\bf p}_b({\bf p};\lambda {\bf v})=
{\bf p}  + \lambda {\bf v}( \lambda {\bf v} {\bf p}
\frac{1}{v^0+1} +\epsilon({\bf p})) \eqno{(2.2b)}$$
where $v^0$, that is given by eq.(2.1d), takes \textit{ the same value}
for $\lambda=+1,-1$. 
\vskip 0.5 truecm
\noindent
Being the particle on-shell, the \textit{independent} transformation is that
of the three-momentum ${\bf p}$, displayed in eq.(2.2b).
For the development of the model we highlight the two following properties of
this transformation :
\vskip 0.5truecm
\noindent
(i) the transformation is not linear with respect to
this variable  ${\bf p}$ due to the presence of $\epsilon({\bf p})$ 
in the r.h.s of that equation;
\vskip 0.5 truecm
\noindent
(ii) the transfomation only depends on the product
$\lambda {\bf v}$.  
\vskip 0.5 truecm
\noindent
For the construction of the PF RHD we shall introduce the infinitesimal 
generator of the Lorentz boost, denoted in the present work as $\bf t$.
To this aim one has, first, to expand  eq.(2.2b) up to the first order
in  the physical velocity $\bf u$ and, second, to construct the operator $\bf t$ 
that transforms the momentum eigenstates according to that expansion [1-4].

\vskip 0.5 truecm
\noindent
Furthermore, we point out that we shall construct a \textit{ unitary} representation
of the boost generator in order to represent the Lorentz transformations according to the 
standard rule of quantum mechanics, i.e. analogously to nonrelativistic quantum mechanics.

\noindent
This procedure seems  suitable for the study 
of systems composed of particles
whose (strong) interaction is described by means of a quasipotential operator.

\noindent
On the other hand, we recall that the spinors of the Dirac equation
are transformed by \textit{pseudounitary} operators, that will be shown
explicitly in eq.(2.22).
In the Dirac equation formalism, unitarity is recovered in the context
of  field theory [24], whose construction is beyond the scope of the present work.

\noindent
However, we shall show that the two representations are connected 
by the transformation of eq.(2.18). Finally, we remark that the price paid
for using a unitary representation of the boost operator is the \textit{nonlocal}
character of both the Hamiltonian operator and of the boost generator, 
that will be explicitly defined in eqs. (2.4 b-e) for positive energy states
and, in general, in eqs.(2.10) and (2.11).

\vskip 0.5 truecm 
\noindent
We now introduce for a single particle 
all the 10 infinitesimal generators of the Poincar\'e group, collectively denoted as
 $[g^I]$ (I=1,...,10). In more detail, these generators are: 
the 3-momentum $\bf p$, i.e. the generator of the spatial translation;
the angular momentum $\bf j$ , i.e. the generator of the spatial rotations;
$\bf t$, i.e. the generator of the Lorentz transformations;
finally, the Hamiltonian $h$,
i.e. the generator of  the time translations.
Their commutation rules are
$$[p^{\alpha},p^{\beta}]= [p^{\alpha},h]=[j^{\alpha},h]=0 \eqno{(2.3a)}$$
$$[j^{\alpha},j^{\beta}]= i \epsilon^{\alpha \beta \gamma}j^{\gamma} \eqno{(2.3b)}$$
$$[j^{\alpha},p^{\beta}]= i \epsilon^{\alpha \beta \gamma}p^{\gamma} \eqno{(2.3c)}$$
$$[j^{\alpha},t^{\beta}]= i \epsilon^{\alpha \beta \gamma}t^{\gamma} \eqno{(2.3d)}$$
$$[t^{\alpha},p^{\beta}]= i \delta^{\alpha \beta} h \eqno{(2.3e)}$$
$$[t^{\alpha},t^{\beta}]= -i \epsilon^{\alpha \beta \gamma}j^{\gamma} \eqno{(2.3f)}$$
$$[t^{\alpha},h]= i p^{\alpha} \eqno{(2.3g)}$$
By introducing the operator $\bf r$ canonically conjugated to $\bf p$,
that is

\noindent
$[p^{\alpha},r^{\beta}]=-i \delta^{\alpha \beta} $,
and the spin operator $\bf s $, the generators $\bf j$, $\bf t$ and $h$ can be put, 
\textit{for the positive energy states} $(\lambda=+1)$,
 in the following form that satisfies the commutation rules of eqs.(2.3a-g).
$${\bf j}= {\bf r} \times {\bf p} + {\bf s} \eqno{(2.4a)}$$
$$h=\epsilon({\bf p}) \eqno{(2.4b)}$$
$${\bf t}= {\bf d} +{\bf g} \eqno{(2.4c)} $$
with
$${\bf d}= \frac{1} {2} ({\bf r}\epsilon({\bf p})+\epsilon({\bf p}){\bf r}) \eqno{(2.4d)} $$
$${\bf g}= \frac{{\bf p} \times {\bf s}} {\epsilon({\bf p}) +m}  \eqno{(2.4e)} $$
As anticipated, the boost is represented by a unitary operator that, for
a finite transformation, has the following form

$$B({\bf v})=\exp(i{\bf t w(v)}) \simeq 1+ i{\bf t u} \eqno{(2.5)}$$
with
$${\bf w(v)}= \frac {{\bf v}} {|{\bf v}|} 
\tanh^{-1}(\frac  {|{\bf v}|} {~v^0})
= \frac {{\bf u}} {|{\bf u}|} 
\tanh^{-1}( {|{\bf u}|} )
  \eqno {(2.6)}$$
Considering a state  
of three-momentum $\bf p$, positive energy $(\lambda=+1)$ 
and z-projection of the spin 
  $\sigma$,
the action of the boost operator on such state is

$$B({\bf v})|{\bf p}, \lambda=+1, \sigma>=
[\frac {\epsilon ({\bf p}_b({\bf p}; {\bf v}))}
{\epsilon ({\bf p})} ]^{1/2} R({\bf p}; {\bf v})
|{\bf p}_b({\bf p}; {\bf v}), \lambda=+1, \sigma > \eqno {(2.7)} $$
\noindent
For these states we adopt the delta normalization shown below in  eq.(2.8).
In consequence,
the numerical factor in the r.h.s. of the previous equation, 
whose origin is due to the nonlinearity
of the Lorentz tranformation with respect to $\bf p$ (see eq.(2.2a)), guarantees 
the correct normalization of the boosted state,
being  $B({\bf v})$ a unitary operator; finally, the rotation operator
$R({\bf p}; {\bf v})$ that is function of the \textit{numerical} vector ${\bf p}$ ,
has been introduced to represent the spin rotation produced by
the operator ${\bf g}$  of eq.(2.4e).
\vskip 0.5 truecm
\noindent
For completeness we also introduce the standard two components spinors  
$w_{\sigma}$ to represent the spin states.
In this way the wave function corresponding $ |{\bf p}, \sigma >$ is written as
$$\psi_{{\bf p} \sigma}({\bf q})=<{\bf q}|{\bf p}, \sigma > = 
w_{\sigma}\delta(\bf q -\bf p) \eqno{(2.8)}$$
and the matrix elements of the boost operator of eq.(2.7) take the form

$$<{\bf q}, \lambda=+1, \mu|B({\bf v})|{\bf p},\lambda=+1, \sigma>=
[\frac {\epsilon ({\bf p}_b({\bf p}; {\bf v}))}
{\epsilon ({\bf p})} ]^{1/2} R_{\mu \sigma}({\bf p}; {\bf v})
\delta({\bf q} - {\bf p}_b({\bf p}; {\bf v})) \eqno{(2.9)}   $$
\noindent
Here $ R_{\mu \sigma}(\bf p; v)$ is the 2 \textsf{x} 2 
matrix representation of the operator
 $R({\bf p};{\bf v})$ acting in the space of the spinors $w_{\sigma}$.
\vskip 0.5 truecm
\noindent
We shall now generalize the procedure outlined above
in order to  include  in the theory also the negative energy states.
These states, as it is shown by the study
of the Dirac equation and by the development of the field theories, 
are introduced for a consistent
relativistic treatment of interacting particles.
To this aim we have to replace, in eq.(2.4b), the positive energy Hamiltonian with
$$h= h(\lambda,{\bf p})=\lambda\epsilon({\bf p}) \eqno{(2.10)}$$
where $\lambda$ represents here an \textit{operator} with the eigenvalues $\lambda=+1$
and $\lambda=-1$
for positive and negative energy, respectively.
In consequence, one can inmediatly verify that the Poincar\'e group
commutation rules of eqs(2.3a-g) can be satisfied
 with the Hamiltonian of eq.(2.10) and by replacing, in eq.(2.4c)
$${\bf t}=\lambda({\bf d} +{\bf g} ) \eqno{(2.11)} $$
By using the previous expression, the generalization of the finite boost for
including the negative energy states is easily found: in eq.(2.5) 
the argument ${\bf v}$ must be replaced by 
$\lambda {\bf v}$, in agreement with the
Lorentz transformation of the three-momentum that was discussed before.  
For the generalized boost operator we shall keep using the notation $B({\bf v})$.

\noindent
In summary, for an Hamiltonian with negative eigenvalues
($\lambda=-1$), also the boost generator must take a minus sign to
give the correct commutation rules. The other generators, $\bf
p$ and $\bf j$ of eq.(2.4a), remain unchanged.
Having introduced for the free particle state
the following ket $|\bf p, \lambda,\sigma>$  
we now need, to represent these states, a \textit{4-component} 
spinorial wave function of the following form 
$$\psi_{{\bf p} \lambda \sigma}^P({\bf q})=<{\bf q}|{\bf p},\lambda, \sigma > =
u^P(\lambda) w_{\sigma}\delta({\bf q} -{\bf p}) \eqno{(2.12)}$$
that satisfies standard ortonormality properties.
We denote this representation of the states as \textit{Poincar\'e
representation}. In the previous equation we have introduced the
following $4 \times 2$ components  \textit{block} spinors
$$u^P(+ )=\left( \matrix {1  \cr
                               0  } \right) \eqno{(2.13a)}$$
$$u^P(- )=\left( \matrix {0  \cr
                               1  } \right) \eqno{(2.13b)}$$
for positive and negative energy, respectively.
These spinors act, in eq.(2.12), on the standard
two component spinor $w_\sigma$.
In the Poincar\'e representation the operator $\lambda$ is a $4\times 4$
\textit{block} matrix of the form
$$\lambda =\left( \matrix {1 & ~0  \cr
                           0 &-1  } \right) \eqno{(2.14)}$$
One straightforwardly obtains the expressions for the Hamiltonian $h$
and for the boost generator $\bf k$ replacing the previous expression
of $\lambda$ in eqs.(2.10) and (2.11).
Boosting the state $|{\bf p},\lambda,\sigma>$ represents a generalization
of eq.(2.7)
$$B({\bf v})|{\bf p},\lambda, \sigma>=
[\frac {\epsilon ({\bf p}_b({\bf p};\lambda {\bf v}))}
{\epsilon ({\bf p})} ]^{1/2} R({\bf p}; \lambda {\bf v})
|{\bf p}_b({\bf p};\lambda {\bf v}),\lambda,\sigma > \eqno{(2.15a)} $$
In consequence, the boosted wave function is of the form

$$\psi_{b; {\bf p} \lambda \sigma}^P({\bf q})=
[\frac {\epsilon ({\bf q})}
{\epsilon ({\bf p})} ]^{1/2}u^P(\lambda) R({\bf p}; \lambda{\bf v})w_{\sigma}
\delta({\bf q} - {\bf p}_b({\bf p};\lambda {\bf v})) \eqno{(2.15b)}   $$
where the equality ${\bf q} = {\bf p}_b({\bf p};\lambda {\bf v})$
given by the delta function has been used in the normalization factor.

\noindent
For the matrix elements of the boost operator, eq.(2.9) is generalized
in the following way
$$<{\bf q}, \lambda', \mu|B({\bf v})|{\bf p},\lambda, \sigma>=
\delta_{\lambda' \lambda}
[\frac {\epsilon ({\bf p}_b({\bf p}; \lambda {\bf v}))}
{\epsilon ({\bf p})} ]^{1/2} R_{\mu \sigma}({\bf p}; \lambda {\bf v})
\delta({\bf q} - {\bf p}_b({\bf p};\lambda {\bf v})) \eqno{(2.15c)}   $$
We now study the connection  of the Poincar\'e representation  with
the one given by the solution of the standard Dirac equation. In
particular this study is very useful in order to construct, in the
following, invariant interaction operators with respect to boost
transformations. We show that the Poincar\'e representation is
completely equivalent to the Dirac one. To this aim we recall that
the solutions of the Dirac equation for a free particle,
 in the momentum space, with the same notations introduced before, 
have the following form

$$\psi_{\lambda {\bf p} \sigma}^D({\bf q})=
u^D(\lambda,{\bf p}) w_{\sigma}\delta({\bf q} -{\bf p}) \eqno{(2.16)}$$
with
$$u^D(+,{\bf p}) = {1\over\sqrt{2\epsilon({\bf p} })}
\left[ \matrix { \sqrt{\epsilon({\bf p}) + m}\cr &\cr
({\bf p} \vec{\sigma} ) \over{\sqrt{\epsilon({\bf p}) +m}}\cr }
\right]\eqno(2.17a)$$

$$u^D(-,{\bf p}) = {1\over\sqrt{2\epsilon({\bf p} })}
\left[ \matrix {- {{({\bf p} \vec\sigma) }
\over{\sqrt{\epsilon({\bf p}) +m}}}\cr &\cr
 \sqrt{\epsilon({\bf p}) +m}\ \cr } \right]\eqno(2.17b)$$

\noindent
In the previous equations the three Pauli matrices $\vec\sigma$ have been introduced.
Note that  in the Dirac representation the spinors  do depend on
$\bf p$ (see the previous equations), while in the Poincar\'e one
( see eqs.(2.13a,b)) they do not.
It is possible to pass from the former representation to the latter by means of
the well-known, unitary but momentum dependent, Foldy-Wouthuysen  (FW)
transformation [25] given 
in the following equation

$$U({\bf p})=\left[\frac {\epsilon({\bf p}) +m}
 {2\epsilon({\bf p}) }\right]^{1/2}+
\left[\frac {\epsilon({\bf p}) -m} 
{2\epsilon({\bf p}) }\right]^{1/2}
\frac  {({\bf p} \vec\gamma )}{|~\bf p|}   \eqno(2.18)$$
where we are introducing the Dirac matrices $\gamma^{\mu}=(\gamma^0,\vec{\gamma})$
in the standard representation.
With straightforward handling one verifies the following properties 
of the FW transformation

$$U^{-1}({\bf p})=U^+({\bf p})=U(-{\bf p}) \eqno(2.19)$$
The transformation for the spinors have the form

$$U({\bf p})u^D(\lambda, {\bf p}) w_{\sigma}= u^P( \lambda) w_{\sigma} \eqno(2.20a)$$
Note that, due to the unitarity of the FW transformation, the spinors 
$u^D(\lambda, {\bf p})$ and $ u^P( \lambda)$
have the same normalization to unity. 
Also, for the complete wave functions, defined in eq.(2.12), one has
the relation

$$U({\bf q})u^D(\lambda, {\bf p}) w_{\sigma}\delta({\bf q} -{\bf p})=
u^P (\lambda) w_{\sigma}\delta({\bf q} -{\bf p}) \eqno(2.20b)$$
Applying the FW transformation as in the previous equation 
to the boosted Poincar\'e wave function given in eq.(2.15b),
the boost transformation of a Dirac wave function
is easily found in the form

$$\psi_{b,\lambda {\bf p} \sigma}^D({\bf q})= 
U^+({\bf q}) \psi_{b,\lambda {\bf p} \sigma}^P({\bf q})=$$
$$=[\frac {\epsilon ({\bf q})}  {\epsilon ({\bf p})}]^{1/2}
u^D(\lambda,{\bf q})
R({\bf p}; \lambda {\bf v})w_{\sigma}
\delta({\bf q} - {\bf p}_b({\bf p};\lambda {\bf v})) \eqno{(2.21)} $$
where, analogously to eq.(2.15b), the equality 
${\bf q} = {\bf p}_b({\bf p};\lambda {\bf v})$
given by the delta function has been used both in the normalization factor
and in the argument of the Dirac spinor.
\vskip 0.5 truecm
\noindent
On the other hand it is well known that in the Dirac theory the spinor boost
is introduced in the form

$$B_D({\bf v})= B_D^+({\bf v})= 
[{1\over 2}(v^0+1)]^{1/2}+
[{1\over 2}(v^0-1)]^{1/2} { ({{\bf v}\gamma^0 \vec\gamma}) \over {| {\bf v}|}}
\simeq $$
$$ \simeq 1+{1\over 2}  ({\bf u} \gamma^0 \vec {\gamma}) \eqno(2.22)$$
where the time component $v^0$ of the four-velocity, 
given in eq.(2.1d), has been used.
Standard calculations show that 

$$B_D({\bf v})u^D(\lambda, {\bf p})w_{\sigma}=
[\frac {\epsilon ({\bf p}_b({\bf p};\lambda {\bf v}))}
{\epsilon ({\bf p})} ]^{1/2} 
u^D(\lambda,{\bf p}_b({\bf p};\lambda {\bf v}))
R({\bf p}; \lambda {\bf v})w_{\sigma} \eqno(2.23)$$
and, consequently, the boosted wave function of eq.(2.21)
can be written as 

$$\psi_{b;\lambda {\bf p} \sigma}^D({\bf q})= 
B_D( {\bf v})u^D(\lambda, {\bf p}) w_{\sigma}
\delta({\bf q} -{\bf p}_b({\bf p};\lambda {\bf v}))\eqno(2.24)$$
For further developments, we note that
eq.(2.23) can be simplified by introducing the \textit{covariantly normalized}
Dirac spinors
$$u^{DC}(\lambda, {\bf p})=[{\epsilon({\bf p})\over m}]^{1/2}
 u^D(\lambda, {\bf p})\eqno(2.25)$$
satisfying the condition $\bar u^{DC}(\lambda, {\bf p})u^{DC}(\lambda, {\bf p})= \lambda$.
The Dirac boost for these spinors is

$$B_D({\bf v})u^{DC}(\lambda, {\bf p})w_{\sigma}=
u^{DC}(\lambda,({\bf p}_b({\bf p};\lambda {\bf v})))
R({\bf p}; \lambda {\bf v})w_{\sigma} \eqno(2.26)$$
We recall that by using Dirac boosts and Dirac matrices one can construct
Lorentz covariant operators.
From standard algebra of the Dirac matrices one has
$$B_D({\bf v})\gamma^0 B_D({\bf v})= \gamma^0 \eqno(2.27a)$$
and
$$B_D({\bf v})\gamma^0 \gamma^{\mu}B_D({\bf v})=
 L^{\mu}_{~\nu}( {\bf v})\gamma^0 \gamma^{\nu}\eqno(2.27b)$$
In consequence, by using eq.(2.26) and eq.(2.27a), one obtains

$$w_{\sigma} ^+ R^+({\bf p}; \lambda {\bf v})
\bar u^{DC}(\lambda,({\bf p}_b({\bf p};\lambda {\bf v})))
u^{DC}(\lambda',({\bf p}_b({\bf p}';\lambda' {\bf v})))
R({\bf p}'; \lambda' {\bf v})w_{\sigma'}=$$

$$ =w_{\sigma}^+ \bar u^{DC}(\lambda,{\bf p})
u^{DC}(\lambda',{\bf p'})w_{\sigma'}  \eqno(2.28a)$$
for  the \textit{scalar} matrix element.
\vskip 0.5 truecm
\noindent
Also, by using eq.(2.26) and eq.(2.27b), one has
$$w_{\sigma} ^+ R^+({\bf p}; \lambda {\bf v})
\bar u^{DC}(\lambda,({\bf p}_b({\bf p};\lambda {\bf v})))
\gamma^{\mu}
u^{DC}(\lambda',({\bf p}_b({\bf p}';\lambda' {\bf v})))
R({\bf p}'; \lambda' {\bf v})w_{\sigma'}=$$
$$ =L^{\mu}_{~\nu}( {\bf v})
w_{\sigma}^+ \bar u^{DC}(\lambda,{\bf p})
\gamma^{\nu}
u^{DC}(\lambda',{\bf p}')w_{\sigma'}   \eqno(2.28b)$$
for the \textit{vector} matrix element.
Similar equations hold for the other Dirac covariants, namely, 
the \textit{pseudoscalar, axial-vector} and \textit{tensor}
matrix elements.

\noindent
Eqs.(2.28a,b), are very important for the construction of invariant
interaction operators that will be done  in the next section.
\vskip 2.5 truecm
\noindent
From eq.(2.26) one obtains the following useful expression
for the spin rotation matrix
$$R({\bf p};  {\bf v})=
\bar u^{DC}(+,{{\bf p}_b}
({\bf p};{ \bf v} ))
B({ \bf v})
u^{DC}(+,{\bf p}) \eqno(2.29)$$
From the previous expression, with standard Dirac algebra
one  also finds the following relation

$$R^+({\bf p};{ \bf v})=R({\bf p}_b ({\bf p};{ \bf v} );- {\bf v})
\eqno(2.30)$$
that will be used in the appendix to show the covariance of the interaction.


\vskip 1.0 truecm

\noindent
\textbf{3. Systems of relativistic interacting particles}
\vskip 0.5 truecm
\noindent
In this section we shall construct a Relativistic Hamiltonian Dynamical 
model for N interacting
spin 1/2 particles also considering negative energy states. 
This objective will be achieved by defining
the 10 total generators of the Poincar\'e group, denoted
with the capital letters
$[G^I]$, in terms of single particle operators. Obviously,
the total generators must satisfy the same  commutation rules 
given in eqs.(2.3a-g) for the single particle generators.
If the interaction were not present, one could easily define the
total generators as the sum of the single particle ones:
$$G^I=\sum_{i=1}^N {g_i}^I \eqno(3.1)$$
automatically satisfying the commutation rules.

\vskip 0.5 truecm
\noindent
As anticipated in the introduction,
the way in which the interaction is introduced makes the difference
among various models of RHD.
For this problem, that has been mainly faced considering 
positive energy states, different solutions have been proposed
as explained in subsect.1.1.
\vskip 0.5 truecm
\noindent
In the present work we want to keep using standard 
Lorentz transformations of the four-vectors 
(canonical boosts), as given in eqs.(2.1 a-c) and (2.2a,b),
so we can only choose IF RHD (i) or PF RHD (ii).

\vskip 0.5 truecm
\noindent
(i) As for the IF RHD,  the interaction 
is added to the sum of the free Hamiltonians but not to the momenta.
In consequence, 
an interaction operator must be also added to the sum of the free
boost operators in order to satify the Poincar\'e algebra commutation 
rules. This method, that is also adopted for the quantization
of the relativistic field theories, has been widely used to introduce 
\textit{relativistic corrections} [2] both to the Hamiltonians 
of bound systems and to the operators that describe the interaction
of these systems with external electromagnetic fields,
significantly improving, for the quark models, the reproduction of the
experimental data [26], specially for the low energy observables.

\noindent
The difficulty of this approach consists in finding the \textit{exact} expression 
of the interaction dependent operator  that modifies the boost generator. 
For this reason we do not follow this method in the present work. 
\vskip 0.5 truecm
\noindent
(ii) In the PF RHD,  the interaction 
modifies both the Hamiltonian and the total momentum of the system,
leaving the boost \textit{ free of the interaction} [5-7].
As shown in the following, the form of the interacting four-momentum operator
can be directly determined.

\noindent
A relevant consequence of the properties of PF RHD  is that it is possible to study 
the dynamics of the composite system in terms of
\textit{explicitly covariant} integro-differencial
wave equations that will be derived in the next section.

\vskip 0.5 truecm
\noindent
We revise the Point Form procedure considering the possibility of
introducing  also negative energy states.

\noindent
Given a (bound) system of N interacting 
constituent particles,
it is possible to observe this system both in its rest reference frame (RF)
and in a generic reference frame (GF).
It is convenient to introduce the
observable quantity

$$V^{\mu}=(V^0,\bf V)$$

\noindent
that represents the four-velocity of the  RF
measured from a GF.
It means that $\bf V$ is the 
parameter that, inserted in eqs.(2.2a,b), 
allows to transform the momenta observed
in the RF of the system to the corresponding quantities
in the GF.  The relation between $V^0$ and $\bf V$ is the same as in eq.(2.1d).
Furthermore, for a system of mass $M$, considering $V^\mu$
as a classical quantity, one has
$$V^{\mu}=({E\over M}, {{\bf P} \over M} )\eqno(3.2)$$
where $E=\sqrt{{\bf P}^2+M^2}$ and $\bf P$ respectively represent 
its energy and  three-momentum  measured in a GF.

\noindent
The procedure to construct the generators of the Poincar\'e group
requires to define $\bf V$ as an \textit{operator}, 
that is as a dynamical variable of the system. This definition will be given in
eq.(3.7).
As first step we introduce the RF 
four-momentum of the i-th particle
$$ p_i^{* \mu}(\lambda_i)=(\lambda_i\epsilon({\bf p}_i^*), {\bf p}_i^*) \eqno(3.3)$$
where the asterisk denotes the quantities observed in the RF.
The sum of these four-momenta over the N constituents,  
by definition of the RF (that is also called \textit{zero momentum} frame),
is given by the following equation

$${\sum_{i=1}^{N}}  p_i^{* \mu}(\lambda_i)=
( {\sum_{i=1}^{N}}  \lambda_i\epsilon({\bf p}_i^*)=M_F, {\bf 0}) \eqno(3.4)$$
where we have also introduced $M_F$ that represents the \textit{free mass operator} 
of the system.
By applying the Lorentz transformation of eq.(2.2b) 
as function of the parameter $\bf V$ 
to the $ p_i^{* \mu}(\lambda_i)$,
also using eq.(3.4), one can write the sum 
of the four-momenta of the particles
in a GF as

$$\sum_{i=1}^N p_i^{ \mu}(\lambda_i)= V^{\mu}M_F \eqno(3.5) $$
\noindent
If $M_F$ is \textit{nonvanishing}
one can solve the previous 
equation with respect to $V^{\mu}$; then by writing
$M_F$ in terms of the $p_i^{ \mu}(\lambda_i)$,
one can express $V^{\mu}$ as
a function the $p_i^{ \mu}(\lambda_i)$, or, more precisely, of the
${\bf p}_i$ and $\lambda_i$, that are chosen as dynamical variables
of the relativistic model.

\noindent
The condition $M_F\neq 0$ is fulfilled by the states in which 
all the particles have 
the \textit{same} energy sign, that is $\lambda_i=\Lambda$
for $i=1,...,N$. These states will be denoted as 
\textit{dynamical states}.
On the other hand, the states in which the particles do not have
all the same energy sign can give a vanishing value of $M_F$, 
not allowing for a definition of $V^{\mu}$ 
in terms of the particle momenta. 
For this reason, these states will be
treated separately as \textit{auxiliary states}.

\noindent
For the dynamical states the expression of $M_F$ as a function 
of the momenta in a GF is the following

$$M_F=M_F(\Lambda,\{{\bf p}\})=
\Lambda \left[ 
\sum_{i,j=1}^N p_i^{ \mu}(\Lambda) 
p_j^{ \nu}(\Lambda) g_{\mu \nu}
        \right]^{1\over 2 } \eqno(3.6)$$
where we have introduced the collective shorthand notation  
$\{{\bf p} \}= {\bf p}_1,...,{\bf p}_N$.
Analogously we introduce the notation  
$\{ \lambda \} =\lambda_1,...,\lambda_N$.
For the dynamical states one has
$\{ \lambda \} =\lambda_1=\lambda_2=...=\lambda_N=\Lambda$.
In consequence,we can also write

$$V^\mu(\{\lambda\}=\Lambda,\{ {\bf p} \})=
 [M_F(\Lambda,\{{\bf p}\})]^{-1}
\sum_{i=1}^N  p_i^{ \mu}(\Lambda)
\eqno(3.7)$$
Let us note that the observable four-vector 
$ V^\mu$, as given in the previous expression,
for both $\Lambda=+1$ and $\Lambda=-1$,
transforms in the same way as a positive energy four-momentum, that is
replacing $\bf p$ with $\bf V$, $\epsilon({\bf p})$ with $V^0({\bf V})$
and setting $\lambda=+1$ in eq.(2.2b). In this way we introduce
$$V^0_b=V^0_b({\bf V};{\bf v})=V^0({\bf V}_b({\bf V} ;{ \bf v}))\eqno(3.8a)$$
$${\bf V}_b= {\bf V}_b({\bf V} ;{\bf v})\eqno(3.8b)$$ 
This result, that is consistent with the definitions of
eqs.(3.2a-c), can be easily derived by transforming,
with the help of eq.(2.2b),
the $ p_i^{ \mu}(\Lambda) $ that appear in eq.(3.7).
\vskip 0.5 truecm
\noindent
We  shall now  choose the complete set of commuting operators that will be used
for the quantum-mechanical description of the system . 
To this aim we note that,
due to its definition in eq.(3.7), 
the operator $V^\mu$ commutes
with the momenta of all the particles. In consequence, 
it is possible to choose the following operators:
the three-momenta of $N-1$ particles 
${\bf p}_1,..., {\bf p}_{N-1} = \{{\bf q} \}$,
the spatial components of the four-velocity $\bf V$,
the energy signs $\{ \lambda \} $, 
and, finally, the spin projections on the z axis
$\sigma_1,...,\sigma_N = \{ \sigma \}$.

\noindent
With this set of commuting operators,
the representation states that will be used to write down 
the wave functions of the model, are of the form
$$ |\psi_r>=  | \{{\bf q} \}, {\bf V} ,  \{ \lambda \}, \{ \sigma \} >
\eqno(3.9)$$
Their normalization is
$$<\psi_r|\psi_r'>=
< \{ {\bf q} \}, {\bf V} ,  \{ \lambda \}, \{ \sigma \}
| \{ {\bf q}'\}, {\bf V}',  \{ \lambda'\}, \{ \sigma'\} >=$$
$$=\delta^3  ({\bf p}_1-{\bf p}'_1)
...\delta^3 ({\bf p}_{N-1}-{\bf p}'_{N-1})
\delta^3({\bf V}-{\bf V}')
\delta_{\{\sigma \} \{ \sigma'\} }
\delta_{\{\lambda \} \{ \lambda'\} }
\eqno(3.10)$$
We choose these representation states in order to derive in a
simple way the \textit{manifestly covariant} wave equation of the model and,
in turn, to obtain a dynamically conserved current, as it will be studied 
in subsequent works.

\vskip 0.5 truecm
\noindent
A different type of representation states, currently denoted as
\textit{velocity states} can be advantageously used to study 
the relativistic bound state wave functions. In the velocity states
the spatial variables are represented by $\bf V $ and by the N (not
indepedendent) rest frame momenta $\{{\bf p}^*\}$ or better by the N-1
(independent) Jacobi momenta $\{\bf k\}$. As shown in ref. [5],
the Lorentz transformation  of these states is given by the standard boost
of $\bf V$, as in eq.(3.8b), and by a Wigner rotation for the 
$\{{\bf p}^*\}$ or for the  $\{\bf k\}$.
\vskip 0.5 truecm
\noindent
In general, 
we point out that, in  our relativistic model, as it will be shown 
in the following, the total three-momentum
$\bf P$  is interaction dependent,  
so it does not commute with the three-momenta 
of the constituent particles
${\bf p}_i$. For this reason
${\bf P}$ cannot be diagonalized simultaneously with them and,
as discussed before, $\bf V$ is conveniently chosen from the beginning.

\vskip 0.5 truecm
\noindent
We note that
the momentum of the $N^{th}$ particle, when it appears
in the calculations, can be expressed  as function of
$ \{ {\bf q} \}, {\bf V} $ and $\{ \lambda \}$. To this aim, we firstly
introduce the four-vector
$$Q^\mu ={\sum_{i=1}^{N-1}} p_i^{\mu}(\lambda_i) \eqno(3.11)$$
\noindent
We also recall the standard relation
$$\lambda_i\epsilon({\bf p}_i^*)=V_\mu p_i^{ \mu}(\lambda_i)\eqno(3.12)$$
Then, we write eq.(3.5) with the definition of $M_F$ given in eq.(3.4) in the form
$$Q^\mu = -p_N^\mu(\lambda_N) + 
V^\mu {\sum_{i=1}^{N}} \lambda_i\epsilon({\bf p}_i^*)\eqno(3.13a) $$

\noindent
Squaring both sides, with the help of eq. (3.12), we find
$$\epsilon({ {\bf p}_N}^*)=
\epsilon_N^*(\{ \lambda \},\{ {\bf q} \}, {\bf V}  )=
\left [ (Q^\mu V_\mu)^2 +m^2- Q^\mu Q_\mu 
\right ]^{1\over 2}\eqno(3.13b) $$
Then
$$ M_F(\{ \lambda \},\{ {\bf q}\}, {\bf V} )=
V^\mu Q_\mu + 
\lambda_N \epsilon_N^*(\{ \lambda \},\{{\bf q} \}, {\bf V} )\eqno(3.13c)$$
and finally, by means of eq.(3.5)
$$ p_N^{\mu}( \{ \lambda \},\{ {\bf q} \}, {\bf V} )=
- Q^\mu + V^\mu   M_F(\{ \lambda \},\{ {\bf q} \}, {\bf V} )\eqno(3.13d) $$
The previous expression can be also used for the auxiliary states.
In fact, when $M_F(\{ \lambda \},\{ {\bf q} \}, {\bf V} )=0$,
one has 
$$ p_N^{\mu}( \{ \lambda \},\{ {\bf q} \}, {\bf V} )=-Q^\mu    $$
In the case of dynamical states, for the developments of the next section 
it is convenient to introduce
$$ M_F( \Lambda =-1,\{ {\bf q} \}, {\bf V})=
  -M_F( \Lambda =+1,\{ {\bf q} \}, {\bf V})=
 -\bar M_F( \{ {\bf q} \}, {\bf V})\eqno(3.13e)$$
Explicit expressions for the \textit{positive} free mass 
$\bar M_F( \{ {\bf q} \}, {\bf V})$ will be given 
in eqs.(4.8) and (4.16a,b) for the two and 
three-body case, respectively.

\noindent
We can now take advantage of eq.(3.2) to define the total momentum 
\textit{operator} as
$$P^{\mu}=M V^\mu \eqno(3.14a)$$
Here $M$ represents the \textit{invariant} mass operator 
of the model, defined as
$$M= M_F(\{ \lambda \},\{ {\bf q} \}, {\bf V} ) +W \eqno(3.14b)$$
where  $W$ represents the Lorentz invariant interaction operator.
Our procedure  for introducing the interaction represents the generalization
of the Bakamjian-Thomas construction [5,27] to a  theory 
with negative energy states. 

\noindent
In order to obtain a Lorentz invariant operator, we require
the following commutation rule of $W$ with the boost generator
$$[W,{\bf T}]=0 \eqno(3.15)$$
We also require
$$[W,{\bf V}]=0 \eqno(3.16)$$
From the previous equation, recalling that 
$V^0= [{\bf V}^2+1]^{1/2}$
one straightforwardly has
$$[W, V^0]=0 \eqno(3.17)$$
ensuring that no commutation problem arises when defining the
total momentum $P^\mu$ in eq.(3.14a).
Furthermore, from the definition of eq.(3.14a) and the requirement
of eqs.(3.16) and (3.17) one has
$$[ P^\mu, M]=0\eqno(3.18)$$
ensuring the space and time translational invariance of the model.

\noindent
In order to satify eqs.(3.15) and (3.16) we choose the interaction 
operator in the form

$$ W=\sum_{ \{ \lambda \} \{ \lambda ' \}}
        W_{ \{ \lambda \} \{ \lambda ' \}} \eqno(3.19a)$$
with
$$  W_{ \{ \lambda \} \{ \lambda ' \}}
=\sum_{ \{ \sigma  \} \{ \sigma '  \}}
\int d^3 \{ {\bf q} \} d^3 \{ {\bf q}'\}d^3{\bf V} ~ 
F (\{ {\bf q} \}, \{ {\bf q}' \})$$

$$ <  \{ {\bf q} \},{\bf V},\{ \lambda\}, \{ \sigma \}  |W^C| 
\{ {\bf q}' \},{\bf V},\{ \lambda ' \},\{ \sigma ' \}>$$

$$ | \{ {\bf q} \}, {\bf V} ,  \{ \lambda  \}, \{ \sigma \} >
   < \{ {\bf q}'\}, {\bf V} ,  \{ \lambda' \}, \{ \sigma' \} |
      \eqno(3.19b)    $$
where the integration $d^3 \{ {\bf q} \}$ simbolizes the integration 
over the momenta ${\bf p}_1$,...., ${\bf p}_{N-1}$ and analogously
for the primed variables;
$F (\{ {\bf q} \}, \{ {\bf q}' \})$ is a function of the momenta, 
that will be explicitly given in eq.(3.20), that ensures the Lorentz covariance 
of the total operator. Also, we introduce the \textit{ manifestly covariant
interaction amplitude} , in the form
$$ <  \{ {\bf q} \},{\bf V},\{ \lambda \},\{ \sigma \}  |W^C |
\{ {\bf q}' \},{\bf V},\{ \lambda '\},\{ \sigma ' \}>=$$

$$\sum_K \sum_{ i>j=1}^N V_{i j}^K
(\{ {\bf q} \}, \{ {\bf q}' \},
\{\lambda \},\{\lambda ' \}, {\bf V})$$
$${w^+}_{\{\sigma \}} \bar u^{DC}(\{ \lambda\},\{ {\bf p}\})
\Gamma_{ij}^K  u^{DC}(\{ \lambda'\},\{ {\bf p}' \})
w_{\{\sigma' \}} \eqno(3.19c) $$
\noindent
in the previous equation
$$V_{i j}^K (\{ {\bf q} \}, \{ {\bf q}' \},
\{\lambda \},\{\lambda ' \},  {\bf V})$$
is a \textit{Lorentz invariant} function, i.e. depending on
the scalar products of the four-momenta,
that expresses the spatial part of the two-body interaction of the model;

\noindent  
$\Gamma_{ij}^K$  can represent:

\noindent
a) the  product of the
covariant Dirac matrices (introduced in the previous section) for the   
particle i and j; more explicitly,
for  $K=1$ one has a \textit{scalar} interaction with
${ I}_i { I}_j$ , being $I_i$ the identity  Dirac matrix 
of the $i^{th}$ particle;
for $K=2$ one has a \textit{ vector} interaction with
$\gamma_i^\nu \gamma_j^\mu g_{\mu \nu}$ and so on for the 
\textit{pseudoscalar, axial-vector}
and \textit{tensor} interactions; but also, for $K > 5$

\noindent
b) terms containing Lorentz invariant products 
of the Dirac matrices 
with the four-momenta of the  $ N-1 $ particles, 
or with $V^\mu$, that is $V_\mu \gamma_i^\mu$,
as we shall see in the next
section for a specific model. 

\noindent
Furthermore, the notation 
$u^{DC}(\{ \lambda\},\{ {\bf p} \})w_{\{\sigma \}} $
represents  the direct product of the Dirac spinors (see eq.(2.25)), 
for all the particles. 
Here and in the following, $p_N^{\mu}$ and the corresponding
primed quantity  are
given by eq.(3.13d).
\vskip 0.5 truecm
\noindent
The structure of eq.(3.19b) immediately shows that the interaction
operator $W$ satisfies the commutation rule of eq.(3.16).
In fact, the interaction amplitude of eq.(3.19c) depends on ${\bf V}$
but the operator   $W$ of eq.(3.19b) has vanishing matrix elements
between states with different values of ${\bf V}$.
Furthermore, the Lorentz invariance of the interaction, that is 
the commutation of the interaction operator with the boost generator,
expressed by eq.(3.15),
is ensured by the invariance of the Dirac spinor matrix elements 
that appear in eq.(3.19c)
taking, in eq.(3.19b),  the function
$F(\{ {\bf q} \}, \{ {\bf q}' \})$
in the form

$$F(\{ {\bf q} \}, \{ {\bf q}' \})=
[\epsilon(\{ {\bf q} \})\epsilon(\{ {\bf q}' \}) ]^{-1/2}
\eqno(3.20)$$
The details of the demonstration are given in the appendix.
According to eq.(A.2), here and in the following subsection ( in particular when 
considering the covariant wave equations) the three-momentum of 
each particle must be transformed according to its energy sign as in eq.(2.2b).

\vskip 0.5 truecm
\noindent
We have now all the elements to prove that  the generators  $[G^I]$
of our relativistic model satisfy the
Poincare algebra commutation rules of eqs.(2.3a-f).

\noindent
The only generators that
contain the interaction are $P^0=H$ and ${\bf P}$ defined by means of eqs.(3.14a,b).
The other ones, being free of the interaction, are given by the sum of 
the single particle generators. For this reason eqs.(2.3b),(2.3d) and (2.3f)
are automatically satisfied.

\noindent
From the definition of  $(P^0=H,{\bf P})$ given in eqs.(3.14a,b)
one immediately obtains the first two relations of eq.(2.3a).

\noindent
From the rotational invariance of $V^0$, $M_F$ and $W$, one obtains the last relation
of eq.(2.3a).

\noindent
Taking also into account the vector character of ${\bf V}$, one verifies eq.(2.3c).

\noindent
Finally, considering the definition of ${\bf V}$ in terms of single particle 
operators as given in eq.(3.7), with $[{\bf T},M_F]=0$ and eq.(A.1)
for the interaction term, one simply
derives eqs.(2.3e,g) completing the verification of the Poincar\'e invariance 
of the model. 


\vskip 1.0 truecm
\noindent
\textbf{4. The wave equation of the model}
\vskip 0.5 truecm
\noindent
In this section we study explicitly the eigenvalue wave equation 
of the model according to the properties discussed in the previous section.
In the subsect.4.1 we shall analyze the general structure of the equation,
higlighting its \textit{manifest covariance},
while in the subsect.4.2,
by using some techniques developed by RWE, we shall introduce the 
\textit{auxiliary states} and the \textit{quadratic terms} of the quasipotential
for two and three-body systems.

\vskip 0.5 truecm
\noindent
\textbf{4.1 General structure of the wave equation }
\vskip 0.5 truecm
\noindent 
The mass eigenvalue equation, by means of  the  mass operator definition of eq.(3.14b),
may be written in  the general form
$$D(M,\{{\bf q} \},\Lambda, {\bf V} )|\Psi>=W |\Psi> \eqno(4.1a)$$
with
$$D(M,\{{\bf q} \},\Lambda, {\bf V} )=
M- M_F(\{{\bf q} \},\Lambda, {\bf V} )\eqno(4.1b)$$
Here we use the free mass operator $M_F$ introduced in eq.(3.13c) 
only for the dynamical states; 
the structure of the interaction operator  $W$  has been given
in eqs.(3.19a-c) and (3.20); the operator
$D(M,\{{\bf q} \},\Lambda, {\bf V} )$
has been introduced here only to simplify the comparison with RWE models.

\noindent

\vskip 0.5 truecm
\noindent
Considering only \textit{dynamical states}, one has

$$ |\Psi>=|\Psi, \Lambda=+1, {\bf V} > + |\Psi,\Lambda=-1, {\bf V} > \eqno(4.2)$$

\noindent
Projecting eq.(4.1) onto the states defined in eq.(3.9)
gives the following set of coupled equations

$$[M- \Lambda \bar M_F (\{ {\bf q} \}, {\bf V} ) ]
\Psi(\{ {\bf q} \},{\bf V},\Lambda,\{ \sigma \})=$$
$$\sum_{\Lambda ' \{\sigma ' \} }
\int d^3\{ {\bf q}' \} <  \{ {\bf q} \},{\bf V},\Lambda,\{ \sigma \}  |W| 
\{ {\bf q}' \},{\bf V},\Lambda ',\{ \sigma ' \}>
\Psi(\{ {\bf q}' \},{\bf V},\Lambda',\{ \sigma ' \})
 \eqno(4.3)$$
where also the definition of the \textit{positive} free mass operator 
given in eq.(3.13e) has been used.

\vskip 0.5 truecm
\noindent
In order to find a \textit{numerical solution}, the previous equation
can be conveniently written in the RF of the bound system, 
then the obtained wave function  is  standardly boosted 
to any GF.

\noindent
In the RF  the sum of the $N$ three-momenta of the particles gives zero, 
as shown in eq.(3.4). This allows to introduce the $N-1$ independent Jacobi momenta
$\{ {\bf k} \}$
that are used to study in a clear way the symmetries of the spatial part 
of the wave function. In this way eq.(4.3) takes the following form
(for brevity, here and in the remainder of the paper, we do not write the RF 
eigenvalue ${\bf V}=0$):

$$[M- \Lambda \bar M_F (\{ {\bf k} \}  ) ]
\Psi(\{ {\bf k} \},\Lambda,\{ \sigma \})=$$
$$\sum_{\Lambda ' \{\sigma ' \} }
\int d^3\{{\bf k}' \} <  \{ {\bf k} \},\Lambda,\{ \sigma \}  |W| 
\{ {\bf k}' \},\Lambda ',\{ \sigma ' \}>
\Psi(\{ {\bf k}' \},\Lambda',\{ \sigma ' \})
 \eqno(4.4)$$

\vskip 0.5 truecm
\noindent
We now turn to write  eq.(4.3) in a
\textit{ manifestly covariant} way.
By introducing
$$\Phi(\{ {\bf q}\},{\bf V},\Lambda,\{ \sigma \})=
[\epsilon(\{ {\bf q} \})]^{1/2}
\Psi(\{ {\bf q} \},{\bf V},\Lambda,\{ \sigma \}) \eqno(4.5)$$
and with the definition of the \textit{manifestly covariant interaction amplitude}
of eq.(3.19c), eq.(4.3) is written as
$$[M- \Lambda \bar M_F (\{ {\bf q} \}, {\bf V} ) ]
\Phi(\{ {\bf q} \},{\bf V},\Lambda,\{ \sigma \})=$$
$$\sum_{\Lambda ' \{\sigma ' \} }
\int { d^3 \{ {\bf q}' \} \over {\epsilon(\{ {\bf q}' \}) }}
<  \{ {\bf q}  \},{\bf V},\Lambda,\{ \sigma \}  |W^C | 
\{ {\bf q}' \},{\bf V},\Lambda ',\{ \sigma ' \}>$$
$$\Phi(\{ {\bf q}' \},{\bf V},\Lambda',\{ \sigma ' \})
 \eqno(4.6)$$
Note that the integrations over the $N-1$ particle momenta is performed in a 
covariant way by means of the factor $ \epsilon(\{ {\bf q}' \}) $
in the denominator of the r.h.s. of the previous equation. 
A similar structure of covariant integration was firstly used in the 
RWE proposed by Gross [13].

\vskip 0.5 truecm
\noindent
We point out that, as \textit{manifest covariance} explicitly shows, 
PF RHD allows to boost in an unambigous way
the wave function of the model for calculating physical observables.

\vskip 0.5 truecm
\noindent
The covariant interaction amplitude  of eq.(4.6) should be determined in order to 
reproduce, with the best possible approximation, the dynamics of the underlying 
field theory.
In particular, the inclusion of the \textit{z-graphs} 
will be performed in the next subsect.4.2 obtaining the 
amplitudes of $W^C_{eff}$ in eqs.(4.13b)
and (4.17b) for the two and three-body case, respectively.

\vskip 0.5 truecm
\noindent
For completeness, note that
the standard form of PF RHD could be recovered by completely excluding in eq.(4.6)
the negative energy states, that is taking only the first term in eq.(4.2). 
Obviously, this choice does not violate the relativistic invariance of
the model. In this way, as anticipated in the introduction,  
the so called Blackenbecler-Sugar wave equation [9] is obtained
without difficulties and
the Poincar\'e algebra commutation rules
are satisfied by generators that only act onto positive energy states.

\noindent
Another possible choice is to include the states with $\Lambda=-1$ and to use 
an interaction operator without \textit{z-graphs}. It corresponds to the Breit equation.
\vskip 0.5 truecm
\noindent
In order to make a comparison with other relativistic models,
we consider explicitly 
the case of a \textit{two-body system}, that is particularly relevant 
in nuclear physics for the study of the deuteron.

\noindent
In this case the only Jacobi variable is the relative momentum
 ${\bf k}={\bf p}_1^*=-{\bf p}_2^*$ 
and eq.(4.4) for the RF takes the form
$$[M-  2 \Lambda  \epsilon({\bf k} ) ]
\Psi( {\bf k} ,\Lambda,\{ \sigma \})=$$
$$\sum_{\Lambda ' \{\sigma ' \} }
\int d^3 {\bf k}'  <   {\bf k} ,\Lambda,\{ \sigma \}  |W| 
 {\bf k}' ,\Lambda ',\{ \sigma ' \}>
\Psi( {\bf k}' ,\Lambda',\{ \sigma ' \})
 \eqno(4.7a)$$
The \textit{manifestly covariant} form of the previous equation
(to be used in a GF), derived directly 
from eq.(4.6), is
$$[M-   \Lambda  \bar M_F ({\bf p}_1, {\bf V} )  ]
\Phi( {\bf p}_1, {\bf V}, \Lambda,\{ \sigma \})=$$
$$\sum_{\Lambda ' \{\sigma ' \} }
\int {d^3 {{\bf p}_1}'\over{\epsilon({\bf p}_1}')}
  <   {\bf p}_1, {\bf V}, \Lambda,\{ \sigma \}  |W^C| 
 {{\bf p}_1}' ,{\bf V},\Lambda ',\{ \sigma ' \}>
\Phi( {{\bf p}_1}',{\bf V},\Lambda',\{ \sigma ' \})
 \eqno(4.7b)$$
where, by means of eqs.(3.13a-e), we can express the positive free mass operator  as
$$\bar M_F ({\bf p}_1, {\bf V} )= 2 p_1^\mu V_\mu =
 2\epsilon( {\bf k}) \eqno(4.8)$$

\noindent
In eqs.(4.7a,b) the two particles have the same energy sign
as in the Breit  equation 
that is obtained by reducing to a three-dimensional form the Bethe-Salpeter equation 
in the case of a \textit{static} interaction between the particles. In particular,
in ref. [14], the Coulomb interaction was considered  to derive  the Breit wave equation
in the RF of the bound system.

\noindent
To reproduce that result in our model,
the  following covariant interaction  amplitude must be inserted in eq.(4.7b) 

$$ <   {\bf p}_1 ,{\bf V}, \{ \lambda\}, \{ \sigma \}  |W^C_{Br}| 
 {\bf p}_1' ,{\bf V},\{ \lambda '\} ,\{ \sigma ' \}>=$$
$$C_{Br}(\lambda_1,\lambda_2, \lambda_1', \lambda_2')
U_{Br}({\bf V},{\bf p}_1, {\bf p}_1') $$
$${w^+}_{\sigma_1} \bar u^{DC}( \lambda_1, {\bf p}_1)
  {w^+}_{\sigma_2} \bar u^{DC}( \lambda_2, {\bf p}_2)$$
$$(V_\mu \gamma_1^\mu ) (  V_\nu \gamma_2^\nu ) 
u^{DC}(\lambda_1', {\bf p}_1') w_{\sigma_1'}
u^{DC}(\lambda_2', {\bf p}_2') w_{\sigma_2'}\eqno(4.9a)  $$
with 
$$C_{Br}(\{\lambda\},\{\lambda'\})= \lambda_1
\delta_{\lambda_1 \lambda_2} \delta_{\lambda_1' \lambda_2'}\eqno(4.9b)  $$
and the covariant function

$$U_{Br}({\bf V},{\bf p}_1,{\bf p}_1',\{\lambda\},\{\lambda'\})=
-m^2{e^2 \over{2 \pi^2}}
[ |V_\mu p_1^\mu(\lambda_1)~ V_\nu p_1{'~^\nu}(\lambda_1') | ]^{-1/2}$$
$$\left[ [V_\mu(p_1^\mu(\lambda_1)- p_1{'~^\mu}(\lambda_1') )]^2-
(p_1^\mu(\lambda_1)- p_1{'~^\mu}(\lambda_1') )
(p_1^\nu(\lambda_1)- p_1{'~^\nu}(\lambda_1') )g_{\mu \nu}
\right]^{-1}\eqno(4.9c) $$
In the RF the two-body Breit equation is obtained in the standard form
of eq.(4.7a) with
$$ <   {\bf k} ,\Lambda,\{ \sigma \}  |W_{Br}| 
 {\bf k}' ,\Lambda ',\{ \sigma ' \}>=  $$
$$-m^2{e^2 \over{2 \pi^2}}C_{Br}(\{\lambda\},\{\lambda'\})$$
$$w^+_{\sigma_1}u^{D+}(\lambda_1 ,{\bf k})w^+_{\sigma_2}u^{D+}(\lambda_2 ,-{\bf k})
 {1\over{({\bf k} - {\bf k}')^2}}
u^D(\lambda_1 ' ,{\bf k}')w_{\sigma_1 '}u^D(\lambda_2' ,-{\bf k}')w_{\sigma_2 '}
\eqno(4.10)$$ 
\vskip 0.5 truecm
\noindent
Summarizing, we
note that our procedure based on the commutation rules of the Poincare algebra,
naturally introduces the \textit{covariant} general expressions of eqs.(4.6);
eqs.(4.7a,b) are simply obtained by specializing eq.(4.6)
for a two-body system.
On the other hand,  the form of the interaction  
is not determined by Poincar\'e algebra and
must be chosen according to
a specific dynamical model for the bound system.
As an example, in the case of the Breit equation discussed above,
the instantaneous approximation is performed
to sum up in a three-dimensional form  the ladder,
or uncrossed, Feynman graphs
introduced in the Bethe-Salpeter equation.
\vskip 0.5 truecm
\noindent 
As for the states with different energy signs, we can analyze in more
detail the point shown in the previous section.
For these states,
the free mass operator $M_F(\{ {\bf q} \},\{ \lambda \}, {\bf V} )$
can be, in general, vanishing and in particular, for two-body systems
with $\{ \lambda \}=( +,-)$ and $\{ \lambda \}= ( -,+) $  
it is \textit{always} vanishing, being 
$\epsilon({\bf p}_1^*)= \epsilon(-{\bf p}_2^*)=\epsilon({\bf k})$.
In consequence,
it is not possible to define, by means of eq.(3.7), the dynamical variable 
${\bf V} $ in terms of the three-momenta of the $N$ particles .
It means  that Poincar\'e invariance, 
in the form discussed in sect.3, only allows to introduce the 
\textit{dynamical states} of eq.(4.2).

\noindent
Such prohibition given by the Poincar\'e algebra naturally leads to the
dynamical consequence of excluding
an unphysical pole at $M=0$ 
in the Green function of the model, 
avoiding the \textit{ continuum dissolution}
or \textit{ cockroach nest}  disease [11,12].
\vskip 0.5 truecm
\noindent
\textbf{4.2 Auxiliary states and quadratic terms}
\vskip 0.5 truecm
\noindent
We first consider the two-body case (\textit{i}), then we generalize the model to the
three-body case (\textit{ii}).

\vskip 0.5 truecm
\noindent
(\textit{i}) Two-body case.

\noindent
The difficulty of the $M=0$ pole was overcome in the Mandelzweig-Wallace (MW) model,
by performing, for two particle bound systems, 
a three-dimen\-sio\-nal reduction of crossed 
and uncrossed Feynman  photon exchange graphs. This procedure, also denoted as
\textit{equal time} reduction, making use
of the eikonal approximation for the \textit{z-graphs}, 
is more accurate than the one of the Breit equation
and correctly reproduces the scattering T matrix up to the terms of second
order in the interaction operator [11,15,17].

\noindent
One has
a  wave equation   written by means of the inverse of the Green function 
that in the RF $ ({\bf V}=0) $ takes the form

$$D_{MW}(M, {\bf k} ,\{ \lambda \} )=
\delta_{\lambda_1 \lambda_2}M + d(\lambda_1,\lambda_2)
2\epsilon({\bf k})\eqno(4.11)$$
with $d(+1,+1)=-1$ and $d(+1,-1)=d(-1,+1)=d(-1,-1)=+1$.

\noindent
The MW equation can be used with different kinds of interaction operators.
Without entering here into details, one can use an operator of the same type
as that given in eqs.(4.9a-c), replacing   
$C_{Br}(\{\lambda\},\{\lambda'\})$
with
$$ C_{MW}(\{\lambda\},\{\lambda'\})= 1-2\delta_{\lambda_1 \lambda_2}
\delta_{\lambda_1' \lambda_2'}\delta_{\lambda_1  -1}\delta_{\lambda_1' -1}
\eqno(4.12)$$
allowing for matrix elements with the states that have different energy signs.

\noindent
However, in that model, it is not possible to  write the wave equation
by defining  the   mass \textit{ operator}, 
as required by the Poincar\'e group commutation
rules in the framework of the Relativistic Hamiltonian Dynamics.

\vskip 0.5 truecm
\noindent
Considering as starting point a two-body system, we propose to keep  
the definition of the mass operator given in the present work  
and to use eqs.(4.7a,b) as the dynamical wave equation of the model.
We include the states with different energy signs, that represent the \textit{z-graphs}, 
by means of a re-definition
of the interaction operator and by the explicit introduction of the
\textit{auxiliary states}.

\noindent
We introduce the following
effective  interaction operator that contains the \textit{z-graphs} in 
the second \textit{quadratic} term, in the form
$$<   {\bf k} ,\Lambda,\{ \sigma \}  |W_{eff}| 
 {\bf k}' ,\Lambda ',\{ \sigma ' \}>=$$
  
$$<   {\bf k} ,\Lambda,\{ \sigma \}  |W_{MW}| 
 {\bf k}' ,\Lambda ',\{ \sigma ' \}>+$$ 
$$ \sum_{\{\bar \lambda''\} \{\sigma''\} }\int d^3{\bf k}'' 
<   {\bf k} ,\Lambda,\{ \sigma \}  |W_{MW}|
  {\bf k}'',\{\bar \lambda''\} ,\{\sigma''\} > $$
$$ {1\over{2\epsilon({\bf k}'')}}
< {\bf k}'',\{\bar \lambda''\} ,\{\sigma''\}|W_{MW}|
 {\bf k}' ,\Lambda ',\{ \sigma ' \}> \eqno(4.13a)
$$

\noindent
where, in the last term, the states with mixed energy signs
are represented by $\{\bar \lambda ''\}=(+1,-1)$, $(-1,+1)$.
The corresponding manifestly covariant amplitude is:
$$<   {\bf p}_1, {\bf V},\Lambda,\{ \sigma \}  |W^C_{eff}| 
 {{\bf p}_1}' ,{\bf V},\Lambda ',\{ \sigma ' \}>=$$
  
$$<   {\bf p}_1 ,{\bf V}, \Lambda,\{ \sigma \}  |W^C_{MW}| 
 {{\bf p}_1}' , {\bf V},\Lambda ',\{ \sigma ' \}>+$$ 
$$ \sum_{\{\bar \lambda''\} \{\sigma''\} }
\int {d^3{{\bf p}_1}'' \over  {\epsilon({{\bf p}_1}'')}}
<   {{\bf p}_1} ,{\bf V},\Lambda,\{ \sigma \}  |W^C_{MW}|
  {{\bf p}_1}'', {\bf V},\{\bar \lambda''\} ,\{\sigma''\} > $$
$$[\bar M_F ({{\bf p}_1}'', {\bf V} )]^{-1} 
 < {{\bf p}_1}'', {\bf V}, \{\bar \lambda''\} ,\{\sigma''\}|W^C_{MW}|
 {{\bf p}_1}' ,{\bf V},\Lambda ',\{ \sigma ' \}> \eqno(4.13b)$$
where we have used  the factor of eq.(3.20)
and we have expressed the denominator of eq.(4.13a)  by means of eq.(4.8).

\noindent
For the calculation of the bound state properties,
the effective interaction matrix elements of eqs.(4.13a) and (4.13b)
must be inserted in the eigenvalue equations eqs.(4.7a) and (4.7b), respectively. 

\vskip 0.5 truecm
\noindent
Also, assuming that the states with equal energy signs
represent the \textit{dominant} contributions in the MW equation,
we can define, in our model, the \textit{auxiliary states}  as
$$\Psi( {\bf k} ,\{\bar \lambda \},\{ \sigma \})=$$
$${1\over{2\epsilon({\bf k})}} 
\sum_{\Lambda '  \{ \sigma '\}} \int d^3{\bf k}'
<   {\bf k} ,\{ \bar \lambda \},\{ \sigma \}  |W_{MW}|
 {\bf k}' ,  \Lambda ',\{ \sigma '\}>  
\Psi( {\bf k}' , \Lambda ' ,\{ \sigma ' \})
\eqno(4.14a)$$
The corresponding covariant definition is 

$$\Phi( {\bf p}_1, {\bf V},\{\bar \lambda \},\{ \sigma \})=
[\bar M_F ({{\bf p}_1}, {\bf V} )]^{-1} $$
$$ \sum_{\Lambda '  \{ \sigma '\}} 
\int {d^3{\bf p}_1'\over{\epsilon({\bf p}_1')}}
<   {\bf p}_1,{\bf V},\{ \bar \lambda \},\{ \sigma \}  |W^C_{MW}|
 {\bf p}_1' , {\bf V}, \Lambda ',\{ \sigma '\}>  
\Phi( {\bf p}_1',{\bf V}, \Lambda ' ,\{ \sigma ' \})
\eqno(4.14b)$$

\noindent
Considering a perturbative expansion 
up to the second order in the interaction, one has the \textit{same result}
that is given by the MW equation. In this sense, our model is dynamically equivalent
to that by MW.

\noindent
Furthermore, as explained in subsect.1.2, the explicit introduction of 
the auxiliary states, even if they do not
participate directly in the dynamics, can be very important for the
calculation of the electroweak interaction matrix elements of the bound system.

\vskip 0.5 truecm
\noindent
(\textit{ii}) Three-body case.

\noindent
\vskip 0.5 truecm
\noindent
We  now propose   the structure of a possible generalization of the model
for a bound system with $N =3$, that is relevant for baryonic quark models,
including the states with mixed energy signs. The same procedure can be straightforwardly
generalized to the case of $N> 3$ particles.

\noindent
First, we specialize  eqs.(4.4) and (4.6) to the case of a three-body system, 
obtaining in the RF

$$[M-  \Lambda  \bar M_F ({\bf p}_{\rho}, {\bf p}_{\lambda} )   ]
\Psi(   {\bf p}_{\rho}, {\bf p}_{\lambda}   ,\Lambda,\{ \sigma \})=$$
$$\sum_{\Lambda ' \{\sigma ' \} }
\int d^3   {\bf p}_{\rho}' d^3 {\bf p}_{\lambda}'   
  <    {\bf p}_{\rho}, {\bf p}_{\lambda}   ,\Lambda,\{ \sigma \}  |W| 
 {\bf p}_{\rho}', {\bf p}_{\lambda}'  ,\Lambda ',\{ \sigma ' \}>
 \Psi(   {\bf p}_{\rho}', {\bf p}_{\lambda}' ,\Lambda',\{ \sigma ' \})
 \eqno(4.15a)$$
where  we have introduced the standard RF three-body Jacobi momenta 

\noindent
${\bf p}_{\rho}= \sqrt{1\over 2}({\bf p}_1^* -{\bf p}_2^*)$ and
${\bf p}_{\lambda}= \sqrt{3\over 2}({\bf p}_1^* +{\bf p}_2^*)$.

\noindent
The \textit{manifestly covariant} form of the previous equation
(to be used in a GF), is derived directly 
from eq.(4.6) in the form
$$[M-   \Lambda  \bar M_F ({\bf p}_1,  {\bf p}_2, {\bf V} )  ]
\Phi( {\bf p}_1, {\bf p}_2, {\bf V}, \Lambda,\{ \sigma \})=$$
$$\sum_{\Lambda ' \{\sigma ' \} }
\int {d^3 {{\bf p}_1'}\over{\epsilon({\bf p}_1'})}
     {d^3 {{\bf p}_2'}\over{\epsilon({\bf p}_2'})}
  <   {\bf p}_1,  {\bf p}_2, {\bf V}, \Lambda,\{ \sigma \}  |W^C| 
 {{\bf p}_1'} , {{\bf p}_2'} ,  {\bf V},\Lambda ',\{ \sigma ' \}>$$
$$\Phi( {{\bf p}_1'}, {{\bf p}_2'},  {\bf V},\Lambda',\{ \sigma ' \})
 \eqno(4.15b)$$
The  positive free mass operator in eq.(4.15a) is
$$\bar M_F ({\bf p}_{\rho}, {\bf p}_{\lambda} )   =
\epsilon({\bf p}_1^*)+\epsilon({\bf p}_2^*)+\epsilon({\bf p}_3^*)=$$
$$\left[ \left( \sqrt{1\over 6}{\bf p}_{\lambda}+
 \sqrt{1\over 2}{\bf p}_{\rho}
\right)^2 +m^2\right]^{1/2}+
\left[ \left( \sqrt{1\over 6}{\bf p}_{\lambda}-
 \sqrt{1\over 2}{\bf p}_{\rho}
\right)^2 +m^2\right]^{1/2}
 $$
$$+\left[ {2\over 3}  {\bf p}_{\lambda}^2
+m^2\right]^{1/2} \eqno(4.16a)$$
With the help of the eqs.(3.11), (3.12),(3.13b) and (3.13e), 
the positive free mass operator in eq.(4.15b) is
$$\bar M_F ({\bf p}_1,  {\bf p}_2, {\bf V} )  =
(p_1^\mu+p_2^\mu)V_\mu 
+\left[ ((p_1^\mu+p_2^\mu)V_\mu)^2 +m^2 -(p_1^\mu+p_2^\mu)(p_1^\nu+p_2^\nu)g_{\mu \nu}
 \right]^{1/2}
\eqno(4.16b) $$

\noindent
For the effective interaction, in order to take into account the effects
of the states with mixed energy signs, we use the following expression

$$ <   {\bf p}_{\rho}, {\bf p}_{\lambda},\Lambda,\{ \sigma \}  |W_{eff}| 
 {\bf p}_{\rho}',{\bf p}_{\lambda}',\Lambda ',\{ \sigma ' \}> =$$
$$ <   {\bf p}_{\rho}, {\bf p}_{\lambda},\Lambda,\{ \sigma \}  |W| 
 {\bf p}_{\rho}',{\bf p}_{\lambda}',\Lambda ',\{ \sigma ' \}> +$$

$$\sum_{\{\bar \lambda''\} \{\sigma''\} }
\int d^3  {\bf p}_{\rho}'' d^3 {\bf p}_{\lambda}'' 
  <   {\bf p}_{\rho}, {\bf p}_{\lambda},\Lambda,\{ \sigma \}  |W| 
 {\bf p}_{\rho}'',{\bf p}_{\lambda}'', \{\bar \lambda''\}  ,\{ \sigma '' \}> $$

$${1\over{ B( {\bf p}_{\rho}'',{\bf p}_{\lambda}'',\{\bar \lambda''\} ) }}
< {\bf p}_{\rho}'',{\bf p}_{\lambda}'', \{\bar \lambda''\}    ,\{ \sigma '' \}|W| 
{\bf p}_{\rho}',{\bf p}_{\lambda}', \Lambda ',\{ \sigma ' \}> \eqno(4.17a)$$  
For the corresponding covariant amplitude we have
$$ <  {\bf p}_1, {\bf p}_2,{\bf V},\Lambda,\{ \sigma \}  |W^C_{eff}| 
 {\bf p}_1', {\bf p}_2',{\bf V},\Lambda ',\{ \sigma ' \}> =$$
$$ <  {\bf p}_1,{\bf p}_2,{\bf V},\Lambda,\{ \sigma \}  |W^C| 
  {\bf p}_1', {\bf p}_2',{\bf V},\Lambda ',\{ \sigma ' \}>  +$$

$$\sum_{\{\bar \lambda''\} \{\sigma''\} }
\int {d^3 {{\bf p}_1''}\over{\epsilon({\bf p}_1''})}
     {d^3 {{\bf p}_2''}\over{\epsilon({\bf p}_2''})}
 <  {\bf p}_1,{\bf p}_2 ,{\bf V},\Lambda,\{ \sigma \}  |W^C| 
 {\bf p}_1'' ,{\bf p}_2'',{\bf V}, \{\bar \lambda''\}    ,\{ \sigma '' \}>$$
$${1\over{ B( {\bf p}_1'',{\bf p}_2'' ,{\bf V}, \{\bar \lambda''\} ) }}
< {\bf p}_1'',{\bf p}_2'' ,{\bf V}, \{\bar \lambda''\}    ,\{ \sigma '' \}|W^C| 
  {\bf p}_1',{\bf p}_2',{\bf V},\Lambda ',\{ \sigma ' \}> \eqno(4.17b)$$  
As in the two-body case the second \textit{quadratic} term represents the
\textit{z-graphs} of the model. 
 
\noindent
The auxiliary states are expressed in general form as

$$\Psi(   {\bf p}_{\rho}, {\bf p}_{\lambda}, \{\bar \lambda \},\{ \sigma \} )= 
{1\over{ B(   {\bf p}_{\rho}, {\bf p}_{\lambda} , \{\bar \lambda\} ) }}$$
$$\sum_{   \Lambda '  \{ \sigma '\}} 
\int d^3 {\bf p}_{\rho}' d^3 {\bf p}_{\lambda}'
 <  {\bf p}_{\rho}, {\bf p}_{\lambda}, \{\bar \lambda \}  ,\{ \sigma \}  |W| 
{\bf p}_{\rho}', {\bf p}_{\lambda}',\Lambda ',\{ \sigma ' \}> 
 \Psi( {\bf p}_{\rho}', {\bf p}_{\lambda}',\Lambda',\{ \sigma '\} )
\eqno(4.18a)$$
\noindent
the covariant form is
$$\Phi( {\bf p}_1,{\bf p}_2 , {\bf V} ,\{\bar \lambda \},\{ \sigma \} )= 
{1\over{ B(  {\bf p}_1,{\bf p}_2 , {\bf V}   , \{\bar \lambda\} ) }}$$
$$\sum_{   \Lambda '  \{ \sigma '\}} 
\int {d^3 {{\bf p}_1'}\over{\epsilon({\bf p}_1'})}
     {d^3 {{\bf p}_2'}\over{\epsilon({\bf p}_2'})}
 <  {\bf p}_1,{\bf p}_2 , \{\bar \lambda \}  ,\{ \sigma \}  |W^C| 
{\bf p}_1',{\bf p}_2',{\bf V},\Lambda ',\{ \sigma ' \}> $$
$$ \Phi( {\bf p}_1',{\bf p}_2' ,{\bf V},\Lambda',\{ \sigma '\} )
\eqno(4.18b)$$

\noindent
where the notation $ \{\bar \lambda \} $ refers, as before, to the
states with mixed energy signs. The specific form of the interaction amplitude
should  be fixed on the base of dynamical considerations that go beyond
the scope of the present work that essentially concerns Poincar\'e invariance.
For the \textit{invariant, nonvanishing} quantity 
$ B({\bf p}_{\rho}, {\bf p}_{\lambda}, \{\bar \lambda\} ) $ or 
$ B( {\bf p}_1,{\bf p}_2 ,{\bf V}, \{\bar \lambda\} ) $ 
in a previous study devoted to three body systems [28], we proposed
a model in which they take the form

$$B({\bf p}_{\rho}, {\bf p}_{\lambda},+1,+1,-1 )=
2\left[ {2\over 3}  {\bf p}_{\lambda}^2+m^2\right]^{1/2} 
\eqno(4.19a)$$

$$B({\bf p}_1,{\bf p}_2,{\bf V},+1,+1,-1)=
2\left[\bar M_F ({\bf p}_1,  {\bf p}_2, {\bf V} )  
-V_\mu ( p_1^\mu +p_2^\mu) \right]
\eqno(4.19b)$$
and

$$B({\bf p}_{\rho}, {\bf p}_{\lambda},+1,-1,-1 )=
\left[ \left( \sqrt{1\over 6}{\bf p}_{\lambda}+
 \sqrt{1\over 2}{\bf p}_{\rho}
\right)^2 +m^2\right]^{1/2}
-\bar M_F ({\bf p}_{\rho}, {\bf p}_{\lambda} )   
\eqno(4.20a)$$

$$B({\bf p}_1,{\bf p}_2,{\bf V},+1,-1,-1)=
  V_\mu p_1^\mu - \bar M_F ({\bf p}_1,  {\bf p}_2, {\bf V} )       \eqno(4.20b)$$

\noindent
The results for other values of $\{ \bar \lambda \}$ are determined by symmetry reasons.

\noindent
This model consists in a generalization of the MW equation to three body
systems that gives the correct nonrelativistic limit.
Further investigations must be developed on the \textit{dynamics} of the
model in order to determine the relevant Feynman graphs (and their
three-dimensional reduction) for a three-body system.
\vskip 0.5 truecm
\noindent
We conclude remarking that Poincar\'e invariance, implemented by means of
the Relativistic Hamiltonian Dynamics, offers a reliable framework to study
in a covariant way the strongly interacting bound systems, also including
negative energy states. The wave functions of the model can be exactly boosted to
any reference frame. 
Manifestly covariant wave equations can be written
avoiding some pathologies encountered in other models.

\noindent
Numerical investigations are required to study the effects of negative energy states
on the electroweak form factors.

\noindent
At a more fundamental level, it would be of great interest to study
in detail the connection between 
the Relativistic Hamiltonian Dynamics and the formalism of
an underlying field theory, that, for the quark models, is assumed to
be the QCD. 

\vskip 1.0 truecm
\noindent
\textbf{ Appendix}
\vskip 0.5 truecm
\noindent
Instead of proving directly eq.(3.15), we study  the finite
boost transformation  
$$B({\bf v})W B^+({\bf v}) = W \eqno(A.1)$$
that, expanded at the first order in ${\bf v}$, is equivalent
to eq.(3.15). Furthermore, in this way we also learn how the quantities
that appear in the model are transformed under finite boost.
The procedure can be summarized in the following steps.
\vskip 0.5 truecm
\noindent
(i) We write down in the next equation
the explicit expression of the action of 
the boost  operator onto the representation states of eq.(3.9). 
By considering
the boost of eq.(2.15a) for a single particle, one obtains for the
total boost
$$B({\bf v})| \{ {\bf q}  \},{\bf V} ,  \{ \lambda  \}, \{ \sigma \} > =
     \left[ {\epsilon(\{ {\bf q}_b \})
 V^0_b
\over  {\epsilon(\{ {\bf q} \} ) V^0 } } 
\right]^{1/2}$$
$$R(\{ {\bf p} \}, \{\lambda \};{\bf v})
|\{ {\bf q}_b \},
{\bf V}_b, \{ \lambda  \}, \{ \sigma \} 
>\eqno(A.2)$$
In the previous expression , 
$ \{ {\bf q}_b \}=  \{ {\bf q}_b \}(\{ {\bf q} \},\{\lambda \}; {\bf v} )$ denotes
the set of the  standardly boosted momenta for the particles $1,...,N-1$
as functions of the unboosted momenta, of the energy signs and of the
parameter of the Lorentz transformation, respectively;
$\epsilon(\{ {\bf q}_b \})$
represents the product of the corresponding boosted absolute values 
of the energies;
recalling that ${\bf V} $ must be considered as a function 
of the particle momenta as shown in eq.(3.7),
when a boost operator is applied to 
eigenstate of ${\bf V}$, one obtains an eigenstate of 
${\bf V}_b={\bf V}_b({\bf V};{\bf v}) $. Its   
 expression  and that of $V^0_b$ have been given in eqs.(3.8a,b);
finally, $R(\{ {\bf p} \},\{\lambda \};{\bf v})$ represents the product 
of the spin rotation matrices for all the particles. 

\vskip 0.5 truecm
\noindent
(ii) We calculate  $B({\bf v})W B^+({\bf v})$ by using the 
operatorial relation of eq.(A.2) onto the interaction defined 
in eqs.(3.19a,b) and then replace the integration variables
$\{ {\bf q} \}$ with  $\{ {\bf q}_b \}$ and analogously
for the primed momenta; we also replace ${\bf V}$ with  ${\bf V}_b$. 
These change of variables introduce the following transformation
factors

$$ d^3 \{ {\bf q} \} =
{\epsilon (\{ {\bf q} \})\over {\epsilon (\{ {\bf q}_b \})}}
~d^3 \{  {\bf q}_b \} \eqno(A.3)$$
and the analogous one for the primed momenta; also
 $$ d^3 {\bf V} = {V^0\over V^0_b}~ d^3 {\bf V}_b \eqno(A.4)$$
\noindent
At this point we re-define  the \textit{integration} variables 
eliminating the index $b$ in 
$ \{ {\bf q_b} \},  \{{\bf q_b}' \}$ and $ {\bf V}_b $,
but now the arguments $ \{ {\bf q} \} , \{ {\bf q}' \}$ and ${\bf V}$
appearing in eq.(3.19b) must be expressed in terms of \textit{inverse}
Lorentz transformations  (obtained with the boost parameter $-{\bf v}$)
of the new integration variables. The result is

$$  B({\bf v}) W_{ \{ \lambda \}\{ \lambda ' \}}B^+({\bf v}) 
=\sum_{ \{ \sigma  \},\{ \sigma '  \}}
\int d^3 \{ {\bf q} \} d^3 \{{\bf q}'\} d^3{\bf V} $$
$$F (\{ {\bf q}_b^- \},
     \{ {\bf q}_b'^-\}   )$$
$$\sum_K \sum_{ i>j=1}^N V_{i j}^K
(\{ {\bf q} \}, \{ {\bf q}' \},
\{ \lambda \},\{ \lambda '\}, {\bf V})$$
$${w^+}_{\{\sigma \}} 
\bar u^{DC}(\{ \lambda \},\{ {\bf p}_b^- \} )
\Gamma_{ij}^K  
     u^{DC}(\{ \lambda'\},\{ {\bf p}_b'^-\} )
w_{\{\sigma' \}} $$
$$R (\{ {\bf p}_b^- \},\{\lambda \};-{\bf v})
 | \{ {\bf q}  \}, {\bf V} ,  \{ \lambda  \}, \{ \sigma \} > $$
 $$  < \{ {\bf q}' \}, {\bf V} ,  \{ \lambda' \}, \{ \sigma' \} |
 R^+(\{ {\bf p}_b'^- \} 
,\{\lambda'\};-{\bf v})$$

$$\left[ {\epsilon(\{ {\bf q}_b^- \})
          \epsilon(\{ {\bf q}_b'^-\}) }
\over  {\epsilon(\{ {\bf q}  \} ) 
        \epsilon(\{ {\bf q}' \} )  } 
\right]^{1/2}
\eqno(A.5)    $$
where the shorthand notation $\{ {\bf p}_b^- \}=\{ {\bf p}_b\} 
( \{ {\bf p}_b \},\{ \lambda\}; -{\bf v} ) $ and the analogous for the
other momentum variables have been introduced.
\vskip 0.5 truecm
\noindent
(iii) By considering the sum over the  complete set of the spin variables, 
the rotation operators can act on the spinors 
$w_{ \{\sigma ' \}}$ and $w^+_{ \{\sigma  \}}$ .
Then one uses  for these operators the property of eq.(2.30) 
with the boost parameter equal to $ -\bf v$ and the transformation
of the Dirac covariant matrix elements of eqs.(2.28a) and (2.28b).

\noindent
Finally, by using the function defined in eq.(3.20)
one verifies eq.(A.1) for each $W_{ \{ \lambda \} \{ \lambda' \}}$

\end{document}